\documentclass[letterpaper]{article} % DO NOT CHANGE THIS
\usepackage{aaai22}  % DO NOT CHANGE THIS
\usepackage{times}  % DO NOT CHANGE THIS
\usepackage{helvet}  % DO NOT CHANGE THIS
\usepackage{courier}  % DO NOT CHANGE THIS
\usepackage[hyphens]{url}  % DO NOT CHANGE THIS
\usepackage{graphicx} % DO NOT CHANGE THIS
\usepackage{natbib}  % DO NOT CHANGE THIS AND DO NOT ADD ANY OPTIONS TO IT
\usepackage{caption} % DO NOT CHANGE THIS AND DO NOT ADD ANY OPTIONS TO IT
\DeclareCaptionStyle{ruled}{labelfont=normalfont,labelsep=colon,strut=off} % DO NOT CHANGE THIS
\usepackage{pifont}
\usepackage{amsmath}
\usepackage{booktabs}
\usepackage{amsfonts}
\usepackage{enumitem}
\usepackage{multirow}
\usepackage{tabulary}
\usepackage{tabularx}
\usepackage{subcaption}
\usepackage{diagbox}

\newcommand{\cmark}{\ding{51}}

\newcommand{\xhdr}[1]{\vspace{0.2mm}\noindent{{\bf #1.}}}

\begin{document}

\title{Taking Advice from (Dis)Similar Machines: \\The Impact of Human-Machine Similarity on Machine-Assisted Decision-Making}

\author {
    % Authors
    Nina Grgi\'{c}-Hla\v{c}a,\textsuperscript{\rm 1,2}
    Claude Castelluccia, \textsuperscript{\rm 3}
    Krishna P. Gummadi \textsuperscript{\rm 1}
}
\affiliations {
    % Affiliations
    \textsuperscript{\rm 1} Max Planck Institute for Software Systems\\
    \textsuperscript{\rm 2} Max Planck Institute for Research on Collective Goods\\
    \textsuperscript{\rm 3} Inria\\
    nghlaca@mpi-sws.org, claude.castelluccia@inria.fr, gummadi@mpi-sws.org
}

\maketitle

% !TEX root = error_types.tex

\begin{abstract}

Machine learning algorithms are increasingly used to assist human decision-making. When the goal of machine assistance is to improve the accuracy of human decisions, it might seem appealing to design ML algorithms that complement human knowledge. While neither the algorithm nor the human are perfectly accurate, one could expect that their complementary expertise might lead to improved outcomes. In this study, we demonstrate that in practice decision aids that are not complementary, but make errors similar to human ones may have their own benefits.

In a series of human-subject experiments with a total of 901 participants, we study how the similarity of human and machine errors influences human perceptions of and interactions with algorithmic decision aids. We find that (i) people perceive more similar decision aids as more \emph{useful}, \emph{accurate}, and \emph{predictable}, and that (ii) people are \emph{more likely to take opposing advice} from more similar decision aids, while (iii) decision aids that are less similar to humans have more opportunities to provide opposing advice, resulting in a higher influence on people’s decisions overall.
\end{abstract}
% !TEX root = error_types.tex

\section{Introduction} \label{sec:intro}

Machine decision aids assist human decision-makers in a variety of scenarios, ranging from medical diagnostics \cite{esteva2017dermatologist} to bail decision-making \cite{propublica_story}. The potential societal impact of using machine decision aids in real-world settings sparked concerns about their accuracy and fairness \cite{propublica_story, barocas_2016}. Decades of research on machine learning can be leveraged to optimize machine decision aids for accuracy, while more recent research in the FAccT community proposed methods for alleviating some concerns about their fairness \cite{dimpact_fpr, friedler_impossibility, grgic2018beyond, hardt2016equality, kleinberg_itcs17, zafar_dmt, zafar_fairness}, accountability and transparency \cite{caruana2015intelligible, datta2016algorithmic, lakkaraju2016interpretable, lakkaraju2017learning, lipton2016mythos, ribeiro2016should, wachter2017counterfactual}.

However, machine decision aids, as the name suggests, do not make the final decisions --- they assist human decision makers. Hence, when designing decision aids, it is crucial to consider not only the decision aids' accuracy and fairness, but also how \emph{human decision-makers} take their advice. 

\begin{figure*}[t]
    \centering
    \begin{subfigure}{.45\textwidth}
        \centering
        \includegraphics[width=0.7\columnwidth]{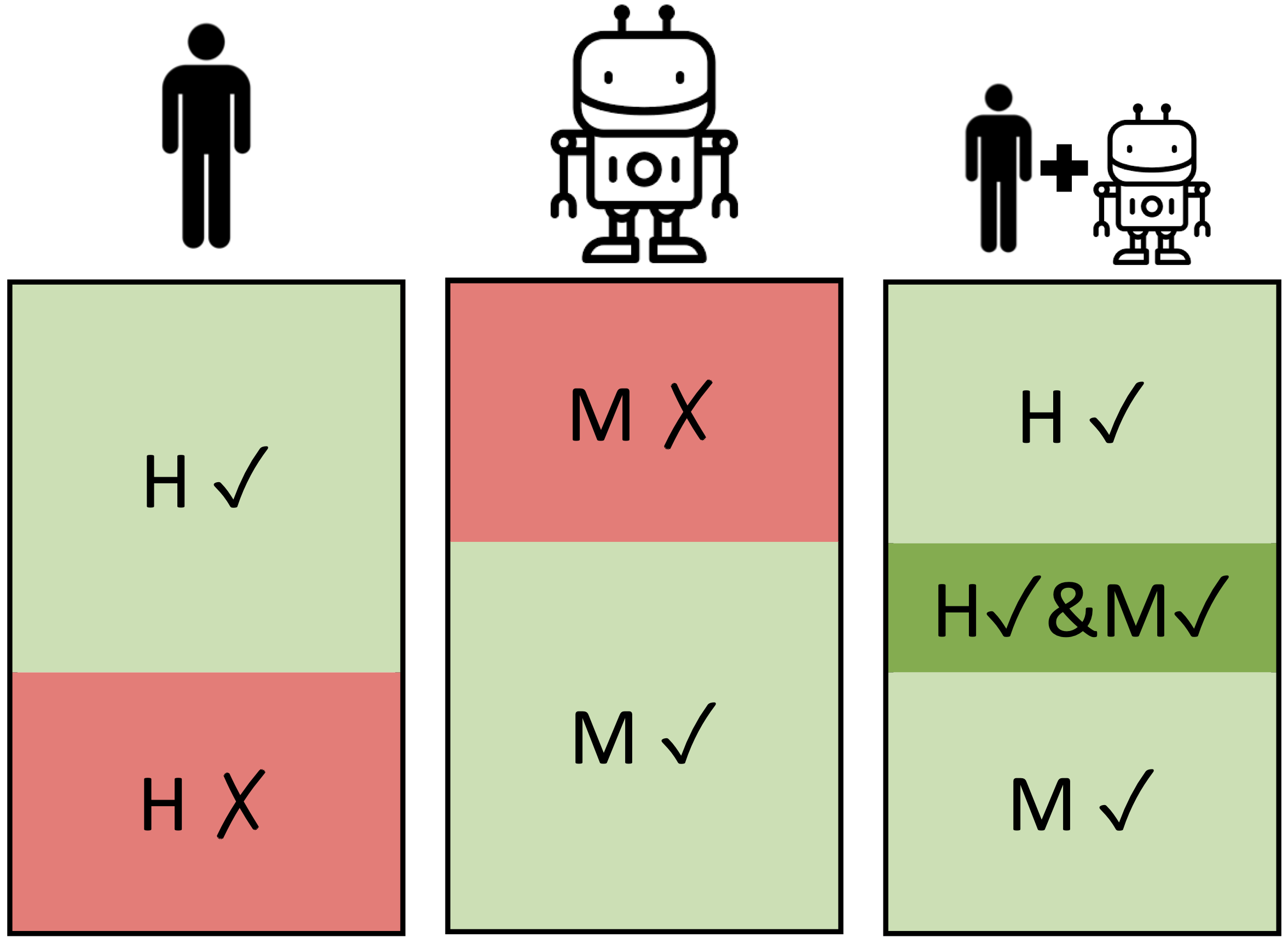}
        \caption{Decision aid with errors \emph{complementary} to human errors.}
        \label{Fig:illustration_complementary}
    \end{subfigure}%
    \hfill
    \begin{subfigure}{.45\textwidth}
        \centering
        \includegraphics[width=0.7\columnwidth]{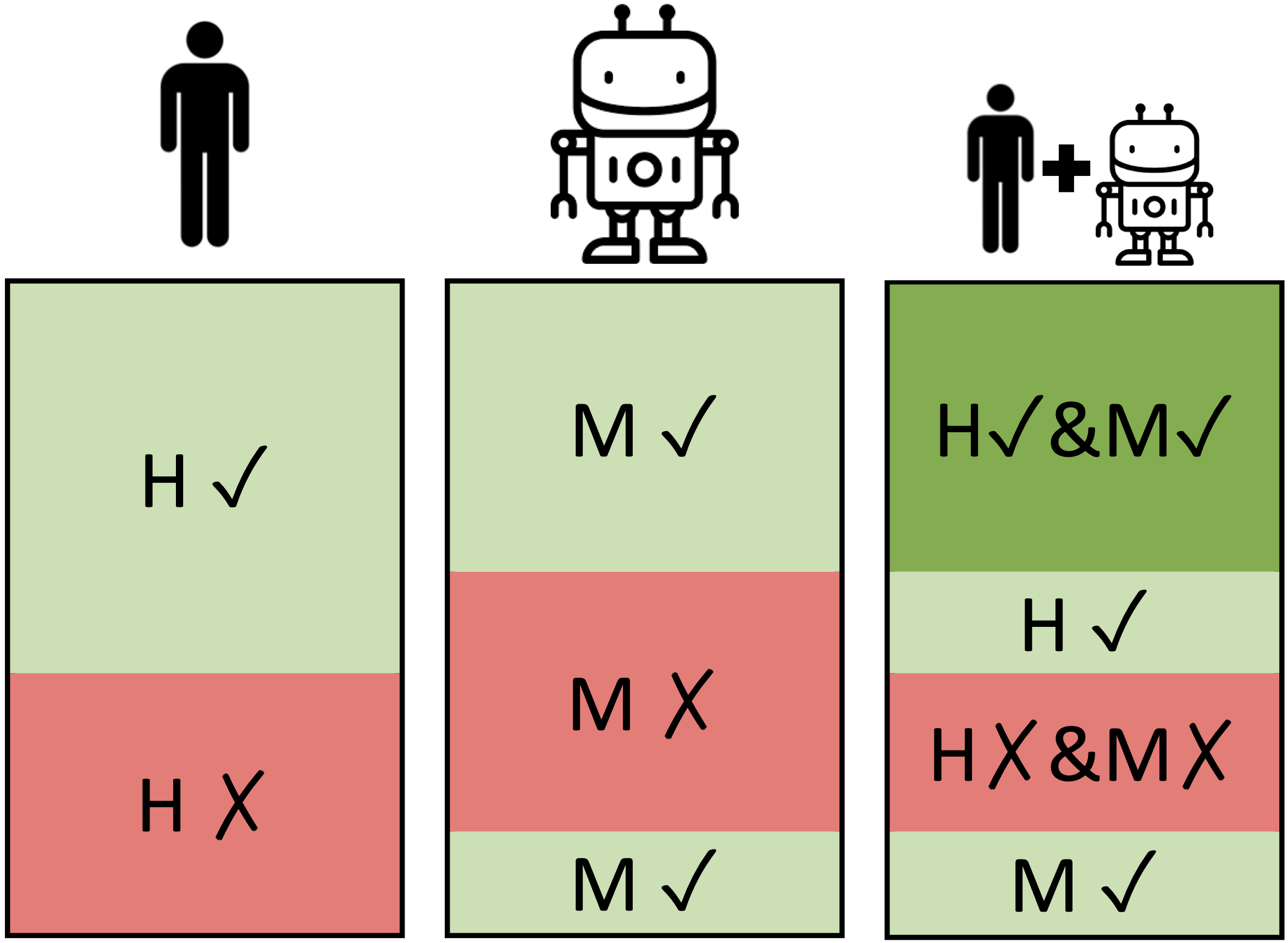}
        \caption{Decision aid with errors \emph{similar} to human errors.}
        \label{Fig:illustration_similar}
    \end{subfigure}%
    \caption{Illustration: Examples of decision aids with errors \emph{complementary} (Fig. \ref{Fig:illustration_complementary}) or \emph{similar} (Fig. \ref{Fig:illustration_similar}) to human errors. Green shading denotes areas where predictions are accurate, while red denotes inaccurate ones. The three panels in both subfigures show the distributions of human, machine, and joint human and machine errors respectively. \emph{Complementary} decision aids provide a better upper bound on the accuracy of joint human and machine decisions (H\cmark $\cup$ M\cmark) in the best case scenario, where humans take all correct advice, but no incorrect advice. On the other hand, \emph{similar} decision aids provide a better lower bound (H\cmark $\cap$ M\cmark) in the worst case scenario, where people take all incorrect machine advice, and no correct advice.}
    \label{Fig:similarity_vs_complementarity}
\end{figure*}

Recent work proposed machine learning algorithms which account for the presence of human agents in their learning procedure \cite{de2020regression, madras2018predict, meresht2020learning, wilder2020learning}. A common thread underlying much of this research is the idea that designing algorithms with skills complementary to human ones may lead to better decision-making outcomes \cite{bansal2019beyond, horvitz2007complementary, kamar2012combining, tan2018investigating, wilder2020learning, zhang2020effect}, in line with the intuition presented in Figure \ref{Fig:similarity_vs_complementarity}. 

How would we expect people to react to advice from a decision aid that complements human skills? Daniel Kahneman's notion of a modern Turing test \cite{kahneman2021neurips} posits that it is acceptable for a system to make mistakes a human might make, but it needs to avoid making mistakes that people would find to be absurd. Additionally, prior research in social psychology found that people are more receptive to advice from advisors more similar to themselves \cite{chan2017effect, faraji2015persuasiveness, suls2000three, yaniv2011receiving}. In this paper, we study the comparative (dis)advantage of using similar or complementary decision aids in machine-assisted decision-making, where human agents make decisions upon receiving machine advice.

\xhdr{Experiment}
In this paper, we experimentally test how the \emph{similarity} of the \emph{decision aid's errors} to typical \emph{human errors} influences human advice-taking behavior. We compare decision aids which are equally accurate overall, but differ with respect to the type and distribution of their errors. Namely, they differ with respect to how similar their errors are to human errors: \emph{human-like}, \emph{anti human-like}, or \emph{random}. Decision aids with \emph{human-like} errors make mistakes for inputs where human respondents are also found to be the least accurate on average. Conversely, decision aids with \emph{anti human-like} errors make mistakes complementary to human ones. They are accurate where humans tend to make mistakes, while making mistakes for inputs where most human respondents make accurate predictions. Finally, decision aids with \emph{random} errors are between these two extremes, and have randomly distributed errors.

I.e., human-like decision aids make mistakes for questions that most people would find difficult to answer correctly, while anti human-like decision aids make mistakes only for questions that the majority of people would find easy to answer correctly. In terms of Kahneman's notion of a modern Turing test \cite{kahneman2021neurips}, errors made by anti human-like decision aids may hence be perceived as egregious or absurd, since most humans would not make such errors.

To quantify the relationship between machine errors and human advice-taking, we conduct a series of human-subject experiments, in which we present respondents with machine advice and measure how the advice influences their decisions. Specifically, we utilize the Judge-Advisor paradigm (JAS), commonly used to study human advice-taking behavior \cite{bonaccio2006advice}, where the decision aids serve as advisors, while human respondents retain decision rights.

To test the robustness of our findings across different domains, we consider three distinct decision-making scenarios: age estimation, criminal recidivism prediction, and dating preference prediction. As an additional robustness check, we consider two degrees of decision aid accuracy: decision aids that achieve the same degree of accuracy as typical human decisions, compared to decision aids that vastly outperform typical humans in terms of decision accuracy.

\xhdr{Contributions}
We conducted a large-scale online study with 901 participants, exploring the impact of machine errors on human advice taking. We found that people's perceptions and advice taking behavior depend on the similarity of the decision aid's errors to typical human errors. In particular, we find that:
\begin{itemize}
    \item People perceive decision aids that make errors similar to human ones as more useful, accurate, and predictable.
    \item People are more likely to take opposing advice from decision aids which make errors more similar to human ones.
    \item Nevertheless, people are significantly more likely to receive \emph{opposing} advice from complementary decision aids. Hence, despite the lower influence of their opposing advice, complementary decision aids have a higher influence on people's decisions overall.
\end{itemize}

We follow up this confirmatory analysis with an exploratory analysis of the effects of human-machine similarity on the \emph{accuracy} of people's decisions. We find that the use of complementary decision aids leads to a slightly greater increase in our respondents' accuracy. However, we also find that complementary decision aids are significantly farther from reaching their full potential for improving accuracy, since people are more likely to take opposing advice from decision aids that are more similar to them.
% !TEX root = error_types.tex

\subsection{Related Work}\label{sec:related_work}
Algorithmic decision aids nowadays advise human decision makers in a plethora of domains, ranging from bail decisions \cite{propublica_story} to medical diagnostics \cite{esteva2017dermatologist}. Hence, it is not surprising that much recent research has studied people's advice taking behavior in machine-assisted decision-making settings. Our research contributes to this interdisciplinary line of work, building up on prior research in social psychology and computer science. 

Research in social psychology has studied how people perceive, react to and utilize machine advice compared to human advice \cite{madhavan2007similarities}. The results are mixed, and the findings vary across decision-making tasks \cite{vodrahalli2021humans}. The majority of studies have reported evidence of algorithm aversion, finding that people tend to favor human advice over machine advice \cite{burton2020systematic, dietvorst2015algorithm, dietvorst2018overcoming, dzindolet2002perceived, mahmud2022influences, prahl2017understanding}. People are also found to perceive machine decisions as less fair and trustworthy than human ones in tasks perceived to require human skills \cite{lee2018understanding}. On the other hand, some studies reported algorithm appreciation \cite{logg2017theory, logg2019algorithm}, finding that people were more receptive to machine advice than to human advice.

The insights related to algorithm aversion from \citet{dietvorst2015algorithm} are particularly relevant for our research: after observing humans and algorithms make the same mistake, people are found to lose confidence in algorithms more quickly than in humans, and opt for human advice over algorithmic advice. In our work, we hypothesize that this preference for human errors over algorithmic errors may go beyond the \emph{identity} of the advisor, and that it may relate to expectations about the advisor's \emph{behavior}. Namely, we study how people react to observing algorithms make errors that are similar to typical human mistakes compared to mistakes that very few people would make.

Recent work in computer science studied how people take advice from machine learning based decision aids. \citet{green2019disparate,green2019principles} studied how machine advice impacts the accuracy and fairness of human decisions. Several studies explored which factors impact the magnitude and quality of a decision aid's influence. For instance, prior work explored the effects of the accuracy of machine advice \cite{salem2015would, yin2019understanding, yu2016trust, yu2017user}, the interpretability and explainability of machine advice \cite{poursabzi2018manipulating, wang2021explanations, zhang2020effect}, providing warnings about machine limitations \cite{engel2021machine}, and varying the stakes associated with the decision-making task \cite{grgic2019human}. 

The work closest to ours is that of \citet{yin2019understanding}, which studied the effects of a decision aid's stated and observed accuracy on human advice taking behavior. In their experiments, they found that people were more likely to take advice from decision aids with a higher stated and observed accuracy than from less accurate ones. In this paper, we hypothesize that people's advice taking behavior depends not only on the \emph{amount} of errors the decision aid makes (i.e., its accuracy), but also on the \emph{type} of errors it makes. Specifically, we explore how human advice-taking behavior is influenced by observing algorithms make errors with varying degrees of similarity to typical human errors.

Much recent research in CS highlights the benefits of human-machine complementarity in joint human-machine decision-making \cite{bansal2019beyond, horvitz2007complementary, kamar2012combining, tan2018investigating, wilder2020learning, zhang2020effect}, especially in settings where algorithms allocate decision rights, such as the learning to defer framework \cite{de2020regression, madras2018predict, meresht2020learning, wilder2020learning}. The intuition is straightforward. During the training phase, the algorithm prioritizes ensuring high predictive accuracy for inputs where humans are expected to make mistakes. Algorithms can make decisions when they are confident in their predictions, and otherwise defer decisions to their human collaborators. However, in many real-world settings algorithms are used as decision aids, whereas humans --- such as judges or doctors --- retain decision rights. Hence, a decision aid's impact depends on people's reactions to the provided advice. In this work, we study if the benefits of complementarity hold in machine-assisted decision-making, and if decision aids more similar to humans also exhibit some desirable properties. 
% !TEX root = error_types.tex

\section{Methodology} \label{sec:methodology}

\xhdr{Hypotheses}
Inspired by the notion of a modern Turing test \cite{kahneman2021neurips} and prior work in social psychology which found that people are more receptive to advice from similar advisors \cite{chan2017effect, faraji2015persuasiveness, suls2000three, yaniv2011receiving}, we build upon prior work in machine-assisted decision-making reviewed in Section \ref{sec:related_work} to form three main hypotheses:

Comparing decision aids of equal accuracy, which differ with respect to the degree of similarity between the decision aid's and typical human errors,

\noindent \textbf{Hypothesis 1:} People perceive similar decision aids as more a) \emph{useful}, b) \emph{accurate}, and c) \emph{predictable}.

\noindent \textbf{Hypothesis 2:} People are \emph{more likely to take opposing advice} from similar decision aids.

Still, since complementary decision aids have less overlap with human decisions, they have more opportunities to give opposing advice. Hence, even though the likelihood of taking any individual piece of opposing advice from complementary decision aids may be lower (H2), we hypothesize:

\noindent \textbf{Hypothesis 3:} Complementary decision aids have a higher \emph{overall influence} on human decisions.

We additionally engage in an exploratory analysis, to investigate the relationship between human-machine similarity and the accuracy of people's decisions.

\subsection{Stimulus Material}
\xhdr{Vignettes}
In our experiments, we consider three different decision-making scenarios: \emph{dating preference} prediction, \emph{criminal recidivism} prediction, and \emph{age} estimation. This set of scenarios covers a broad range of possible applications of machine decision aids. Firstly, these tasks differ with respect to their potential societal impact, with dating preference prediction on one end, and criminal recidivism prediction on the other. Secondly, they differ with respect to the type of thinking required by the decision-maker \cite{daniel2017thinking}, with age estimation being close to System 1, or fast thinking tasks, while dating and recidivism prediction being close to System 2, or slow thinking tasks. Finally, these scenarios provide decision makers with different amounts and types of information (images or natural language text). 

The \emph{dating preference} prediction task leverages data from a speed dating experiment gathered by \citet{fisman2006gender}. After being shown a summary of a speed date, our respondents were asked to predict if the speed dating participant wanted to see their date again. The speed date summaries contained information about the participants' demographics, romantic expectations and impressions about their date, as shown in Figure \ref{Fig:date_description} in the Appendix. \citet{yin2019understanding} previously used this dataset in their study of machine-assisted decision-making, and the format of our vignettes exactly replicates theirs.

The \emph{criminal recidivism} prediction task uses the ProPublica COMPAS dataset \cite{propublica_story}. Our respondents were shown defendants' profiles, and asked to predict if they will commit a new crime within two years. The defendants' profiles contained information about the defendants' gender, age, and criminal history, as shown in Figure \ref{Fig:defendant_description} in the Appendix. This dataset was previously used for studying machine-assisted decision-making \cite{grgic2019human} and algorithmic fairness \cite{dressel2018accuracy, zafar_dmt,grgic2018beyond,corbett2018measure,dimpact_fpr}.

Finally, the \emph{age} estimation task relies on the IMDB-WIKI dataset gathered by \citet{Rothe-ICCVW-2015, Rothe-IJCV-2018}. We showed respondents images of people's faces, and asked them to estimate if they are above or below the age of 21, as depicted in Figure \ref{Fig:age_picture} in the Appendix.

For each of the three datasets, we selected a random subset of 50 data points to use as vignettes in our experiments. All of the datasets were cleaned, and incomplete or otherwise defective data points were removed prior to the random subset selection process. For the age dataset, which was skewed towards images of people above the age of 21, we additionally enforced class balancing constraints.

\xhdr{Decision Aids}
To construct decision aids that have varying degrees of similarity to human errors, we first gathered data about how accurate humans are in our three decision-making scenarios. For each of the scenarios, we recruited approximately 100 respondents (a total of 305 respondents) and asked them to make predictions for all 50 data points from the respective dataset. As a measure of human accuracy for a specific data point, we used the fraction of respondents whose prediction matched the ground truth. For our sample size, the margin of error for human accuracy estimates is $10\%$, for a $95\%$ confidence level.\footnote{After conducting the main experiment, we evaluated whether these human accuracy estimates coincided with the accuracy of people's pre-advice decisions in the main experiment. For each vignette, we calculated the difference between the fraction of people who made an accurate prediction in the first study and in the second study, prior to receiving machine advice. We found that for all three datasets the difference between people's accuracy in the two studies was not significant, with a mean close to 0.}

To construct the decision aids used in the main experiment, we relied on (i) estimates of the fraction of people who make accurate predictions for a given vignette, and (ii) ground truth labels. For each decision aid, we started with the ground truth labels and flipped a certain fraction of labels to achieve the desired degree of human-machine similarity and accuracy.\footnote{We opted for this synthetic approach to generating machine advice in order to ensure that we fully control the structure of the errors, to test our hypotheses. In real-world applications, ML based decision aids would be trained, as discussed in the Design Implications subsection of Section \ref{sec:discussion}.}
Human-like decision aids made errors only where people were most likely to make errors as well. More formally, they maximized the similarity between human and machine decisions, subject to an accuracy constraint. On the other hand, anti human-like decision aids made errors only where people were most likely to make correct decisions. I.e., they made mistakes that people are unlikely to make. More formally, they minimized the similarity between human and machine decisions, subject to an accuracy constraint. Finally, random decision aids had errors distributed uniformly at random, given an accuracy constraint.

As a robustness check, we developed decision aids at two levels of accuracy: human majority vote accuracy and superhuman accuracy. The former set of decision aids achieves the same degree of accuracy as the majority vote of human predictions did for the same decision-making scenario (Age: 76$\%$, COMPAS: 60$\%$, Speed Date: 64$\%$). The latter has a superhuman accuracy of $84\%$, fixed across all three scenarios. We chose an accuracy of $84\%$ to ensure that these decision aids outperform the human majority vote in all decision making tasks, while still making some errors (which is a prerequisite for studying our research questions).

We developed and ran experiments with 18 distinct decision aids ($\{$COMPAS, Age, Speed Date$\}$ $\times$ $\{$human-like, random, anti human-like$\}$ $\times$ $\{$majority vote accuracy, superhuman accuracy$\}$). E.g., for the Age dataset, an anti human-like decision aid with human majority vote accuracy has an accuracy of $76\%$. It makes incorrect predictions for $24\%$ of data points, which were selected according to the accuracy of human predictions. Specifically, since it is the anti human-like decision aid, it makes incorrect predictions for the data points where the observed human accuracy was the highest.
An example of mistakes made by the human-like and anti human-like decision aids for the COMPAS dataset is shown in Figure \ref{Fig:example_mistakes} in the Appendix.

\subsection{Experimental Design}

\xhdr{Experimental Procedure}
Our experiment consisted of two phases: the (first) \emph{test-drive phase} and the (second) \emph{prediction phase}, in line with the design of \citet{yin2019understanding} and \citet{dietvorst2015algorithm}. In each phase, respondents answered questions about 25 vignettes --- a total of 50 vignettes (Figures \ref{Fig:date_description}, \ref{Fig:defendant_description}, and \ref{Fig:age_picture} in the Appendix). For each vignette, respondents made a preliminary pre-advice prediction, before observing machine advice, and then making their final post-advice prediction (Figure \ref{Fig:date_questions} in the Appendix). Before commencing the experiment, participants were shown an introductory text which described the decision-making task and the details of the experimental setup (Figure \ref{Fig:survey_intro} in the Appendix).

In the \emph{test-drive phase}, after each post-advice prediction, respondents received feedback about the accuracy of their own prediction and the decision aid's advice, thereby getting a chance to build a mental model of the decision aid. To incentivize respondents to put effort into building a mental model of the machine's predictions in the test-drive phase of the experiment, we informed them that they could earn monetary rewards in the prediction phase.

In the \emph{prediction phase}, we explored how the mental models formed in the test phase influence people's advice-taking in the second phase. We did not provide feedback about the respondent's and the decision aid's accuracy after every question, in order to minimize the risk of the respondents updating their mental model of the machine's errors (Figure \ref{Fig:date_questions} in the Appendix, without the last paragraph). Following the approach of \citet{dietvorst2015algorithm}, we used monetary incentives only in the prediction phase. For each correct prediction, we rewarded respondents with a $\$0.10$ bonus, and penalized them the same amount for each incorrect prediction. Similar financial incentives have been shown to encourage respondents to provide accurate responses \cite{chittilappilly2016survey,harris2011you}.

At the end of each phase of the experiment, we gathered data about the respondents' perceptions of the decision aid's performance (Figure \ref{Fig:feedback} in the Appendix). Specifically, we asked people to estimate their own and the decision aid's accuracy, and to assess their ability to predict the machine's predictions, as well as the machine's usefulness. We gathered this data twice, once at the end of each phase of the experiment. Immediately afterwards, we informed the participants how well they and the machine decision aid actually performed on the previous 25 questions.

\xhdr{Experimental Conditions}
Our experiment has a between-subjects, randomized, full-factorial design with three factors: human-machine similarity (3 levels), decision-making scenario (3 levels), and decision aid accuracy (2 levels). We employ a repeated measures design, where each respondent makes 50 predictions. 
Each decision-making scenario considers a fixed set of 50 data points. Following the design of \citet{yin2019understanding}, the data points are split into two subsets, ensuring that each decision aid achieves the same predictive accuracy on both subsets respectively. In each phase, we show participants one of the subsets, and vary the order of subsets across participants. To prevent order-bias \cite{redmiles2017summary}, we show the vignettes in random order in each phase. 

\xhdr{Dependent Variables}
We used data about people's pre and post-advice predictions to measure the \emph{influence} of machine advice on peoples \emph{decisions} and their \emph{accuracy}. 
We consider two measures of influence on decisions: (i) \emph{overall} influence, defined as the difference between post-advice and pre-advice agreement with machine advice, and (ii) \emph{conditional} influence, defined as the overall influence for data points where people's pre-advice decisions disagreed with machine advice, i.e., the influence of opposing advice. Similarly, we define the \emph{overall} and \emph{conditional} influence on accuracy by comparing the accuracy of the respondents' pre and post-advice predictions.
We also used data about the decision aid's perceived \emph{usefulness}, \emph{accuracy}, and \emph{predictability}, gathered at the end of each experimental phase.

\subsection{Data Collection}
We recruited respondents using the online crowd-sourcing platform Prolific. Prolific is an alternative to MTurk, commonly used for recruiting participants for online human-subject studies in academic research \cite{palan2018prolific}. We recruited respondents who have self-reported to be US nationals, had an approval rate of at least 95$\%$ on previous studies, and have completed at least 100 studies so far.

In our experiments, we included two simple instructed response items as attention check questions (e.g., "Please respond to this question by selecting \emph{Somewhat disagree} as the answer."). 
Similar instructed response items are commonly employed for identifying careless responses in survey data \cite{meade2012identifying}. In our analysis, we discarded the responses of all respondents who did not complete the survey, or did not successfully complete both attention check questions. Our final sample consists of 901 respondents who successfully answered both attention check questions.

We ran two sets of experiments. In the first experiment we gathered data for modeling the accuracy of human decisions (Section \ref{subsec:modeling_humans}). 305 out of 320 respondents ($95\%$) correctly answered the attention check questions. The average completion time for this set of surveys was 14 minutes, and respondents were paid a base fee of \pounds 2 for taking part in this experiment (i.e., slightly above \$11 per hour).

\begin{table}[t]
\centering
\small
\begin{tabularx}
{\columnwidth}{>{\hsize=1.9\hsize}l|>{\hsize=0.6\hsize}r|>{\hsize=1\hsize}r|>{\hsize=0.5\hsize}r}
\toprule
\textbf{Demographic Attribute} & \textbf{Sec. \ref{subsec:modeling_humans}} & \textbf{Sec. \ref{subsec:perceptions} - \ref{subsec:accuracy}} & \textbf{Census}\\
\hline
Female & 56$\%$ & 48$\%$ & 51$\%$\\
\hline
Asian & 10$\%$ & 14$\%$ & 6$\%$ \\
Black & 7$\%$ & 7$\%$ & 13$\%$ \\
Hispanic & 8$\%$ & 7$\%$ & 18$\%$ \\
White & 70$\%$ & 69$\%$ & 61$\%$\\
Other & 5$\%$ & 3$\%$ & 4$\%$ \\
\hline
$>$ Bachelor's degree & 57$\%$ & 56$\%$ & 30$\%$\\
\hline
$<35$ years & 63$\%$ & 59$\%$ & 46$\%$\\
$35-54$ years & 31$\%$ & 31\% & 26$\%$\\
$55+$ years & 6$\%$ & 10$\%$ & 28$\%$\\
\hline
Liberal & 54$\%$ & 59$\%$ & 33$\%\dagger$\\
Conservative & 12$\%$ & 15$\%$ & 29$\%\dagger$\\
Moderate & 27$\%$ & 23$\%$ & 34$\%\dagger$\\
Other & 7$\%$ & 4$\%$ & 4$\%\dagger$\\
\bottomrule
\end{tabularx}
\caption{Survey Samples. We targeted respondents who have self-reported to be US nationals, and have completed at least 100 studies on Prolific with an approval rate $\geq 95\%$. The respondents' demographics are compared to the 2016 U.S. Census~\citep{census_acs}, and Pew data on political leaning ~\citep{pew_politics} (marked with $\dagger$).}
\label{Tab:Demographics}
\end{table}

\begin{figure*}[t]
    \centering
    \begin{subfigure}{.33\textwidth}
        \centering
        \includegraphics[width=0.99\textwidth]{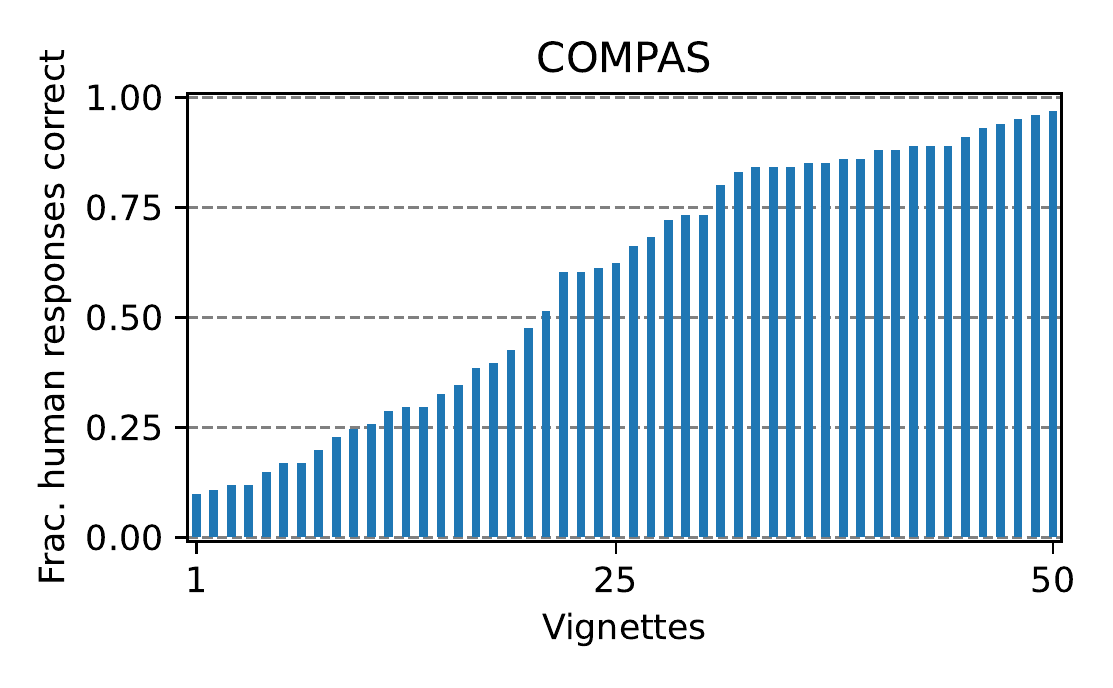}
        \caption{COMPAS}
        \label{Fig:human_errors_COMPAS}
    \end{subfigure}%
    \begin{subfigure}{.33\textwidth}
        \centering
        \includegraphics[width=0.99\textwidth]{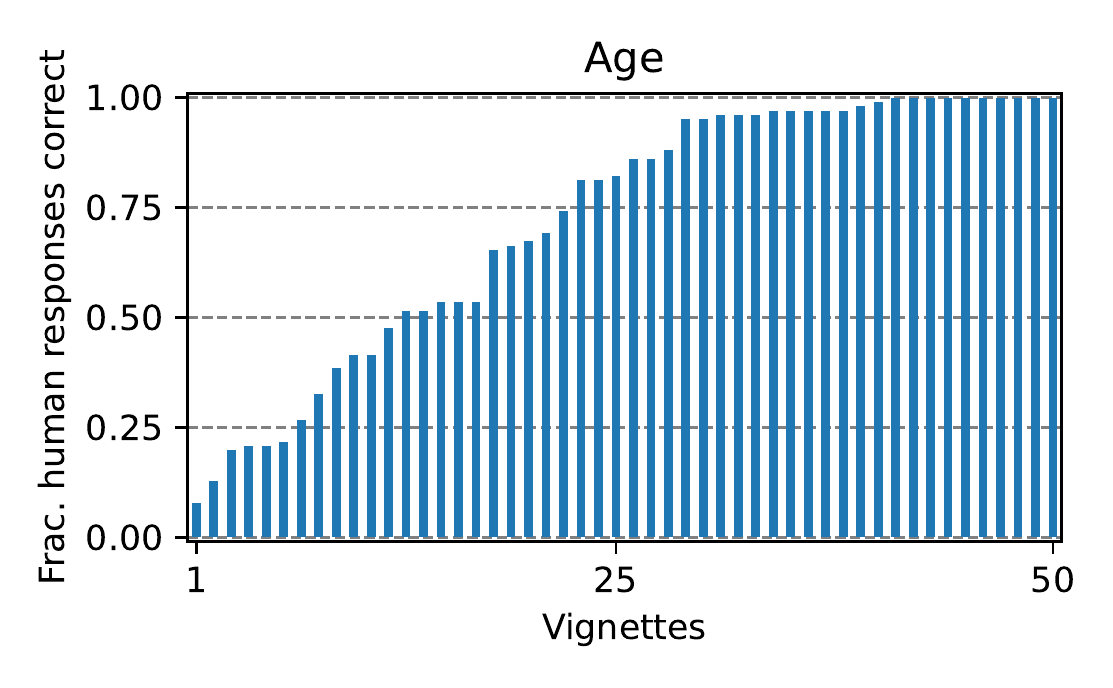}
        \caption{Age}
        \label{Fig:human_errors_age}
    \end{subfigure}%
    \begin{subfigure}{.33\textwidth}
        \centering
        \includegraphics[width=0.99\textwidth]{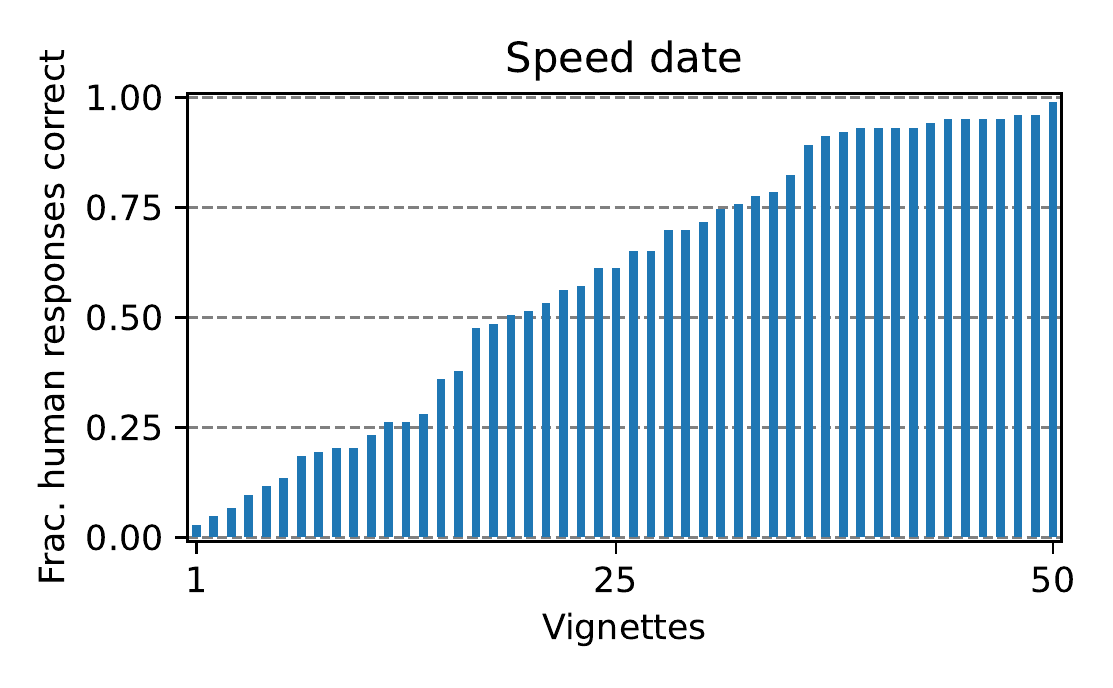}
        \caption{Speed Date}
        \label{Fig:human_errors_speed_date}
    \end{subfigure}%
    \caption{Distribution of human errors for the COMPAS [left], Age [center], and Speed date [right] dataset. The y-axis shows the percent of human predictions that were correct for a given vignette. The vignettes are sorted increasingly w.r.t. y-axis values.}
    \label{Fig:human_errors}
\end{figure*}

In the second experiment we gathered data for testing our hypotheses (Sections \ref{subsec:perceptions} - \ref{subsec:accuracy}). $92\%$ of respondents answered both attention check questions correctly, resulting in a sample of 596 respondents. On average, participants took 21 minutes to complete the survey. Respondents were paid a base fee of \pounds 2.5 for taking part in the experiment (i.e., slightly above \$9.3 per hour). Additionally, the respondents could earn bonus payments based on their performance, as described in Figure \ref{Fig:intro_experiment} in the Appendix.

To assess participant satisfaction and identify any issues with the experiments, participants were asked to answer a series of questions about their experience. In both experiments participants reported that they found the survey interesting (mean ratings of 4.3 and 4.6 on a 5-point Likert scale for the first and second experiment respectively), that they found the questions easy to understand (mean ratings of 4.7 and 4.8 respectively), and that they would like to take part in a similar survey in the future (4.6 and 4.8 respectively).

In Table \ref{Tab:Demographics} we report the detailed demographics of the samples we used to model the accuracy of human decisions (1st column) and to test our hypotheses (2nd column). Since the respondents are US nationals, we additionally compared the sample demographics to the 2016 US Census \citep{census_acs}, and Pew Research Center's data on political leaning \citep{pew_politics} (3rd column). On a high-level, our respondents are more educated, younger, and more liberal than the general US population. Also, our sample consists of more white and Asian respondents, and fewer black and Hispanic respondents than the US population.

\subsection{Analysis}
Throughout the paper, we utilize descriptive statistics to summarize the basic information about our data. We corroborate these findings with statistical hypothesis testing. 

To study the causal effect of the experimental manipulations on people's perceptions and behavior, we employ linear mixed models. We account for repeated measures by including crossed random effects terms for respondents and questions. For H1, we rely on a multivariate multiple linear regression with three dependent variables: perceived usefulness (5-point Likert scale from -2 to 2), accuracy ($[0,1]$) and predictability (5-point Likert scale from -2 to 2). For H2 and H3, we utilize a multiple linear\footnote{We utilize a linear regression for ease of interpretation of coefficients, especially of interaction effects. For a discussion on the applicability of linear models for binary dependent variables, please refer to \citet{hellevik2009linear}. Our results remain qualitatively the same when a logistic regression is applied on the binary dependent variables in H2 and H3.} regression with the overall and conditional influence of advice as dependent variables respectively. For each of the three hypotheses we build two models, one including only human-machine similarity as an independent variable, and one which also includes the control variables (experimental phase, accuracy, and dataset).

Human-machine similarity is a categorical variable with 3 levels: Human-like, Random, and $\neg$Human-like. In all of the models, we use the Random level as the reference category. I.e., the estimated regression coefficients for Human-like and $\neg$Human-like treatments convey information about the effect of the respective treatment compared to the Random treatment. We additionally perform Wald tests on the estimated regression coefficients of the Human-like and $\neg$Human-like treatment to directly compare their effects.
% !TEX root = error_types.tex

\section{Results} \label{sec:results}

% !TEX root = error_types.tex

\subsection{Designing the Decision Aids} \label{subsec:modeling_humans}

The first step in studying the effects of human-machine similarity on people's advice taking behavior was designing machine decision aids with varying degrees of similarity to human errors. To do so, we gathered data about human decisions, to understand for which inputs people typically make accurate predictions, and for which they make mistakes.

In Figure \ref{Fig:human_errors}, we show the distribution of human errors for each of the three datasets separately. For each vignette, we show the percent of respondents who made an accurate prediction (without algorithmic assistance). We find that people's accuracy varies between datasets and between vignettes. Each dataset consists of data points where people are overwhelmingly correct (on the right of the x-axis), and data points where most respondents made incorrect decisions (on the left of the x-axis). In other words, each dataset has vignettes with varying degrees of human accuracy, which will enable us to design decision aids with varying degrees of similarity of human and machine errors.

For the COMPAS dataset, the mean accuracy of people's responses was 0.59. For 60$\%$ of the vignettes, more than half of respondents made an accurate prediction. People's performance in terms of accuracy on the COMPAS dataset was comparable to that reported in prior work \cite{dressel2018accuracy, grgic2019human}. For the Speed date dataset, people had a mean accuracy of 0.59, and an accurate majority vote for 64$\%$ of the vignettes. Finally, for the Age dataset, people were more accurate, with a mean accuracy of 0.72, and an accurate majority vote prediction for 76$\%$ of the vignettes. We used this data to construct human-like (H), random (R) and anti human-like ($\neg$H) decision aids, as described in Section \ref{sec:methodology}.

\begin{figure*}[t]
    \centering
    \begin{subfigure}{.33\textwidth}
        \centering
        \includegraphics[width=0.99\textwidth]{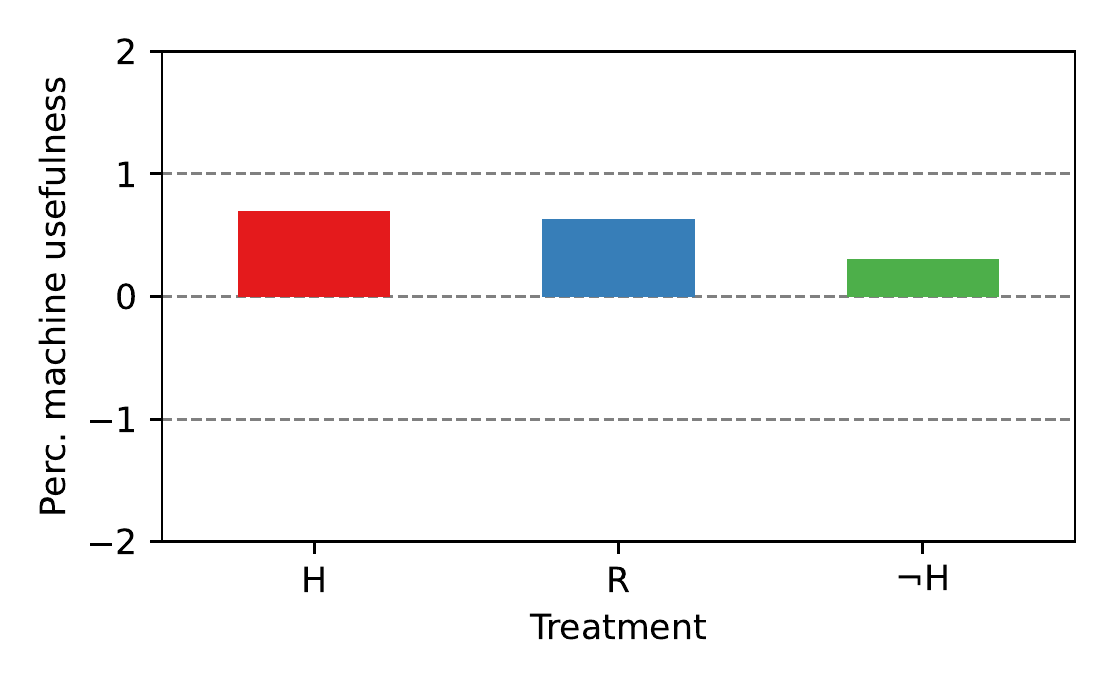}
        \caption{Usefulness}
        \label{Fig:perc_usefulness}
    \end{subfigure}%
    \begin{subfigure}{.33\textwidth}
        \centering
        \includegraphics[width=0.99\textwidth]{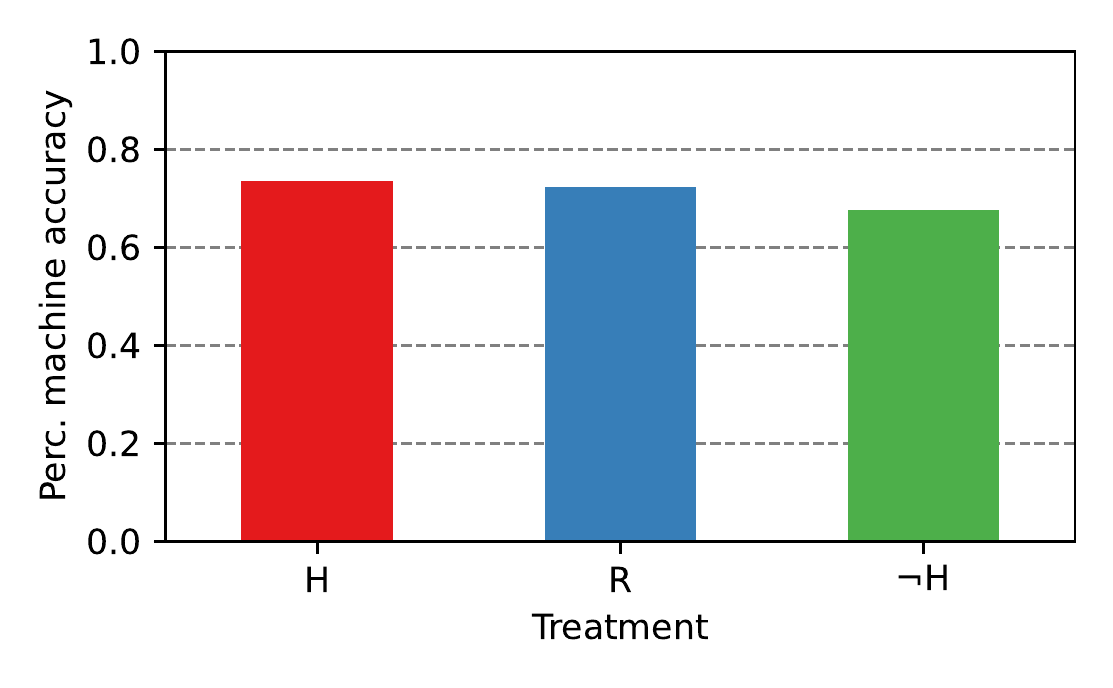}
        \caption{Accuracy}
        \label{Fig:perc_accuracy}
    \end{subfigure}%
    \begin{subfigure}{.33\textwidth}
        \centering
        \includegraphics[width=0.99\textwidth]{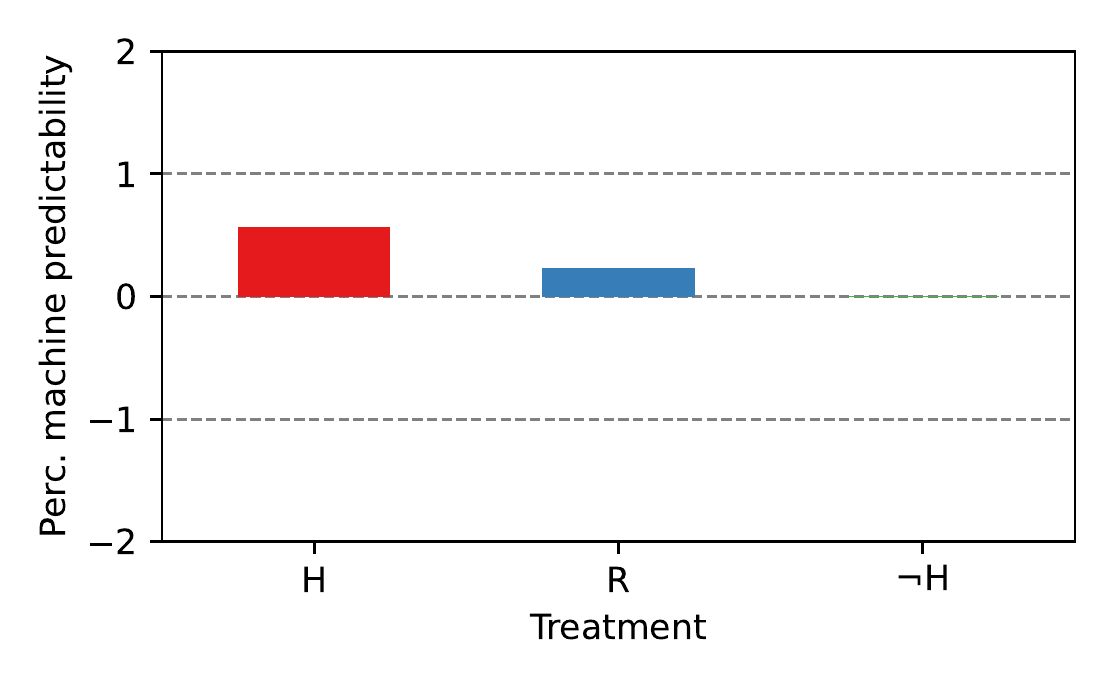}
        \caption{Predictability}
        \label{Fig:perc_predictability}
    \end{subfigure}%
    \vspace{-2mm}
    \caption{\textbf{Perceptions} of machine predictions for decision aids with human-like (H), anti human-like ($\neg$H), and randomly distributed (R) errors. The perceived usefulness and predictability were rated on a 5-point Likert scale. On the plots, -2 corresponds to the lowest rating of usefulness and predictability, and 2 corresponds to the highest. The perceived accuracy was reported as the estimated number of correct predictions, as shown in Figure \ref{Fig:estimate_correctness} in the Appendix, and was converted to a value $\in [0,1]$ in our analysis. Human-like decision aids are perceived as more useful, accurate, and predictable than anti human-like decision aids.}
    \label{Fig:perceptions}
\end{figure*}

\begin{table}[hbt!]
    \centering
    \small
    {\def\sym#1{\ifmmode^{#1}\else\(^{#1}\)\fi}
    \begin{tabulary}{\columnwidth}{L|R|R}
    \toprule
     & \textbf{Model 1} & \textbf{Model 2}\\
    \midrule
    \textbf{Usefulness} & & \\
    Human-like & 0.0674  (0.074)& 0.0500  (0.071)\\
    $\neg$ Human-like & -0.328\sym{***} (0.074)& -0.330\sym{***} (0.071)\\
    Second Phase & &  -0.0419  (0.057)\\
    High Acc. & &  0.673\sym{***} (0.058)\\
    Age D. & &  -0.0351  (0.071)\\
    Speed Date D. & &  0.0199  (0.071)\\
    Intercept & 0.630\sym{***} (0.052)& 0.320\sym{***} (0.076)\\
    \midrule
    \textbf{Accuracy} & & \\
    Human-like & 0.0135  (0.015)& 0.00942  (0.013)\\
    $\neg$ Human-like  & -0.0471\sym{**}  (0.015)& -0.0488\sym{***} (0.013)\\
    Second Phase & &  -0.00450  (0.006)\\
    High Acc. & &  0.145\sym{***} (0.011)\\
    Age D. & &  0.0588\sym{***} (0.013)\\
    Speed Date D. & &  0.00929  (0.013)\\
    Intercept & 0.723\sym{***} (0.011)& 0.630\sym{***} (0.013)\\
    \midrule
    \textbf{Predictability}& & \\
    Human-like & 0.338\sym{***} (0.072)& 0.333\sym{***} (0.071)\\
    $\neg$ Human-like & -0.225\sym{**}  (0.072)& -0.230\sym{**}  (0.071)\\
    Second Phase & &  -0.134\sym{*}  (0.057)\\
    High Acc. & &  0.217\sym{***} (0.058)\\
    Age D. & &  0.197\sym{**}  (0.071)\\
    Speed Date D. & &  -0.0309  (0.071)\\
    Intercept & 0.232\sym{***} (0.051)& 0.136  (0.076)\\
    \bottomrule
    \end{tabulary}}
    \vspace{-2mm}
    \caption{Dependent variables: \textbf{Perceived} usefulness, accuracy and predictability. For non-binary variables, the reference categories are the Random Treatment and the COMPAS Dataset. Standard errors in parentheses. * symbols next to coefficients indicate their statistical significance as follows: * $p<0.05$, ** $p<0.01$, *** $p<0.001$. N = 1192. Wald tests show that the coefficients associated with the human-like and anti human-like treatment are significantly different for all three perceptions, with $p<0.001$.}
    \label{Tab:perceptions}
\end{table}
% !TEX root = error_types.tex

\subsection{Human Perceptions of Machine Performance} \label{subsec:perceptions}

In this section, we present our results on people's perceptions about the decision aids' performance. Specifically, we explore the relationship between human-machine error similarity, and the decision aids' perceived \emph{usefulness}, \emph{accuracy} and \emph{predictability}.

Descriptively, Figure \ref{Fig:perceptions} shows that human-machine similarity is positively correlated with all three perceptions of performance. In Figure \ref{Fig:perc_usefulness}, we see that human-like decision aids are on average rated as more useful than random ones, which are in turn rated as more useful than anti human-like decision aids. The same pattern holds for the perceived accuracy in Figure \ref{Fig:perc_accuracy}, and predictability in Figure \ref{Fig:perc_predictability}. The mean perceived usefulness and predictability of human-like decision aids are 0.4 and 0.6 points higher (on a 5-point Likert scale) than of anti-human like ones, and human-like decision aids are also perceived as 6$\%$ more accurate.

Model 1 in Table \ref{Tab:perceptions} and the subsequent Wald test on its coefficients show that the observed difference between the perceptions of human-like decision aids (H) and anti human-like ones ($\neg$H) is statistically significant with p$<$0.001. However, the difference between the random decision aid (R) and human-like decision aids (H) is significant only for the perceived predictability. 
This gives us
\begin{quote}
    \xhdr{Result 1} People perceive human-like decision aids as more useful, accurate, and predictable than anti human-like decision aids.
\end{quote}

\xhdr{Control Variables}
This pattern holds when we control for the decision-making task, the decision aid's accuracy, and the experimental phase. Figure \ref{Fig:perceptions_separated} in the Appendix shows the results from Figure \ref{Fig:perceptions} separated by these three control variables. Model 2 in Table \ref{Tab:perceptions} shows that the differences between the perceived usefulness, accuracy and predictability of H and $\neg$H are statistically significant even when controlling for the aforementioned variables. 

Model 2 in Table \ref{Tab:perceptions} also provides interesting insights about the effects of the control variables. The accuracy of the decision aid significantly impacted all three perceptions. More accurate decision aids were perceived as more useful, accurate, and predictable. The effects of the experimental phase and the dataset were not consistent across all three perceptions. The experimental phase had a borderline significant effect only on predictability. In the second phase, where people did not observe feedback about the decision aid's performance after every question, people rated decision aids as less predictable. The Speed Date dataset did not lead to significantly different perceptions than the COMPAS dataset. 
However, for the Age dataset, where the baseline human (and hence machine) accuracy was significantly higher than in the other two datasets, people rated the decision aids as more accurate and predictable.

% !TEX root = error_types.tex

\begin{table}[t]
    \centering
    \small
    {
    \begin{tabulary}{\columnwidth}{CC|CCC}
    \toprule
    \multicolumn{2}{c|}{\textbf{Agreement}} & \multicolumn{3}{c}{\textbf{Treatments}}\\
        Pre & Post & H & R & $\neg$H \\
    \midrule
        0 & 0 & 0.145 & 0.306 & 0.418 \\
        0 & 1 & 0.111 & 0.137 & 0.150 \\
        1 & 0 & 0.012 & 0.009 & 0.006 \\
        1 & 1 & 0.732 & 0.549 & 0.426 \\
    \bottomrule
    \end{tabulary}}
    \caption{Distribution of the respondents' decisions with respect to the four possible agreement configurations, for each of the three treatments. The first two columns show the four configurations, where 0 denotes disagreement and 1 denotes agreement with machine advice.}
    \label{Tab:agreement_dist}
\end{table}

\begin{table}[t]
    \centering
    \small
    {\def\sym#1{\ifmmode^{#1}\else\(^{#1}\)\fi}
    \begin{tabulary}{\columnwidth}{L|RR|RR}
    \toprule
    & \multicolumn{2}{c|}{\textbf{Model 1}} & \multicolumn{2}{c}{\textbf{Model 2}}\\
    \midrule
    Human-like & -0.0296\sym{*} & (0.012)& -0.0308\sym{**} & (0.011)\\
    $\neg$ Human-like & 0.0163 & (0.012)& 0.0173 & (0.011)\\
    Second Phase & & & 0.0277\sym{***}& (0.006)\\
    High Acc. & & & 0.0749\sym{***}& (0.010)\\
    Age D. & & & -0.0743\sym{***}& (0.015)\\
    Speed Date D. & & & -0.0104 & (0.016)\\
    Intercept & 0.128\sym{***}& (0.010)& 0.105\sym{***}& (0.013)\\
    \bottomrule
    \end{tabulary}}
    \vspace{-2mm}
    \caption{Dependent variable: \textbf{Overall agreement} change. For non-binary variables, the reference categories are the Random Treatment and the COMPAS Dataset. Standard errors in parentheses. * symbols next to coefficients indicate their statistical significance as follows: * $p<0.05$, ** $p<0.01$, *** $p<0.001$. N = 29800. Wald tests show that the coefficients associated with the human-like and anti human-like treatment are significantly different, with $p<0.001$.}
    \label{Tab:agreement_change}
\end{table}

\begin{table}[t]
    \centering
    \small
    {\def\sym#1{\ifmmode^{#1}\else\(^{#1}\)\fi}
    \begin{tabulary}{\columnwidth}{L|RR|RR}
    \toprule
    & \multicolumn{2}{c|}{\textbf{Model 1}} & \multicolumn{2}{c}{\textbf{Model 2}}\\
    \midrule
    Human-like & 0.124\sym{***}& (0.032)& 0.102\sym{***}& (0.030)\\
    $\neg$ Human-like & -0.0453 & (0.025)& -0.0436 & (0.022)\\
    Second Phase & & & 0.0736\sym{***}& (0.012)\\
    High Acc. & & & 0.184\sym{***}& (0.020)\\
    Age D. & & & -0.111\sym{***}& (0.028)\\
    Speed Date D. & & & -0.0305 & (0.030)\\
    Intercept & 0.309\sym{***}& (0.020)& 0.231\sym{***}& (0.024)\\
    \bottomrule
    \end{tabulary}}
    \vspace{-2mm}
    \caption{Dependent variable: \textbf{Conditional agreement} change, i.e., agreement change for instances where pre-advice decisions disagreed with machine advice. For non-binary variables, the reference categories are the Random Treatment and the COMPAS Dataset. Standard errors in parentheses. * symbols next to coefficients indicate their statistical significance as follows: * $p<0.05$, ** $p<0.01$, *** $p<0.001$. N = 12578. Wald tests show that the coefficients associated with the human-like and anti human-like treatment are significantly different, with $p<0.001$.}
    \label{Tab:agreement_change_conditional}
\end{table}

\begin{figure*}[t]
    \centering
    \begin{subfigure}{.33\textwidth}
        \centering
        \includegraphics[width=0.99\textwidth]{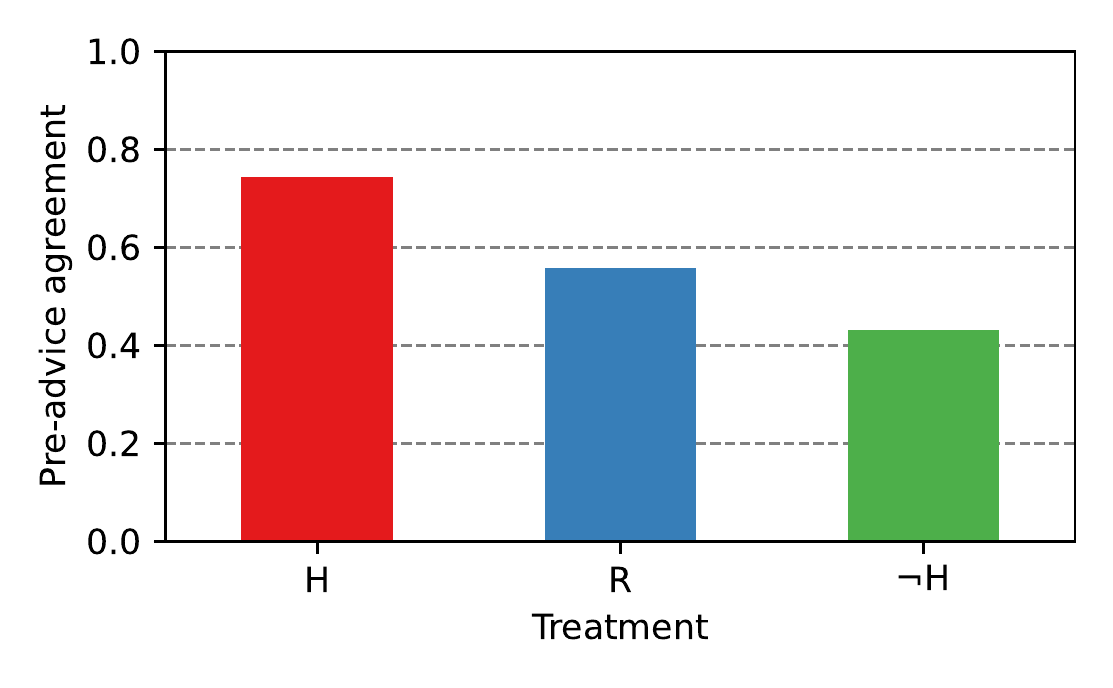}
        \caption{Pre-advice Agreement}
        \label{Fig:agreement_pre}
    \end{subfigure}%
    \begin{subfigure}{.33\textwidth}
        \centering
        \includegraphics[width=0.99\textwidth]{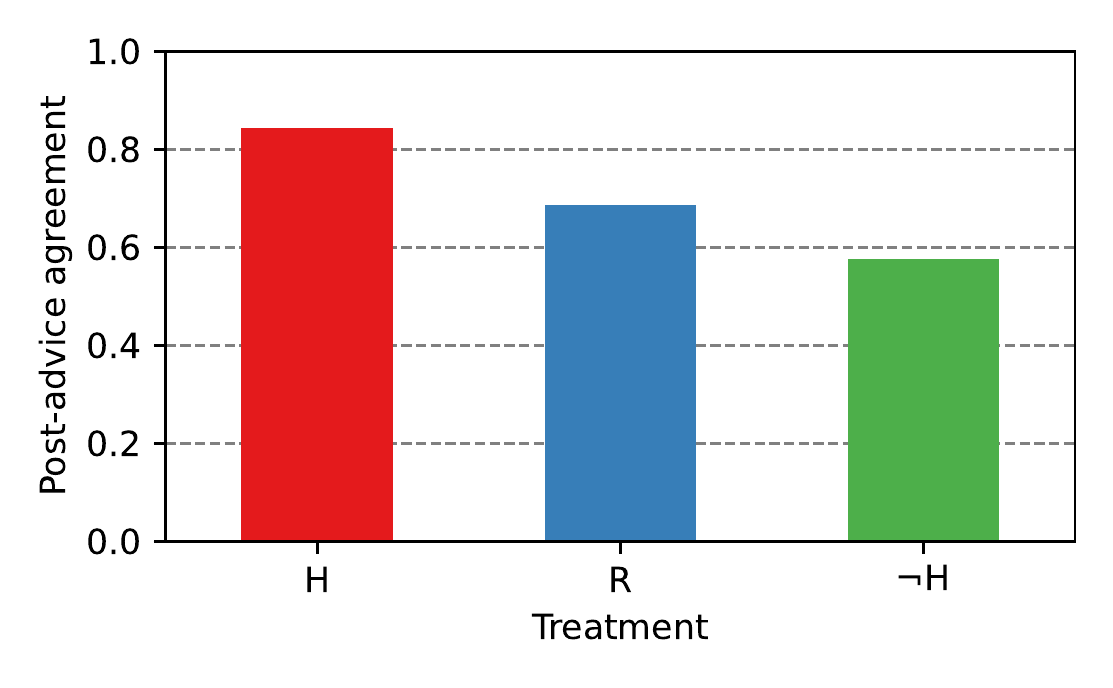}
        \caption{Post-advice Agreement}
        \label{Fig:agreement_post}
    \end{subfigure}%
    \begin{subfigure}{.33\textwidth}
        \centering
        \includegraphics[width=0.99\textwidth]{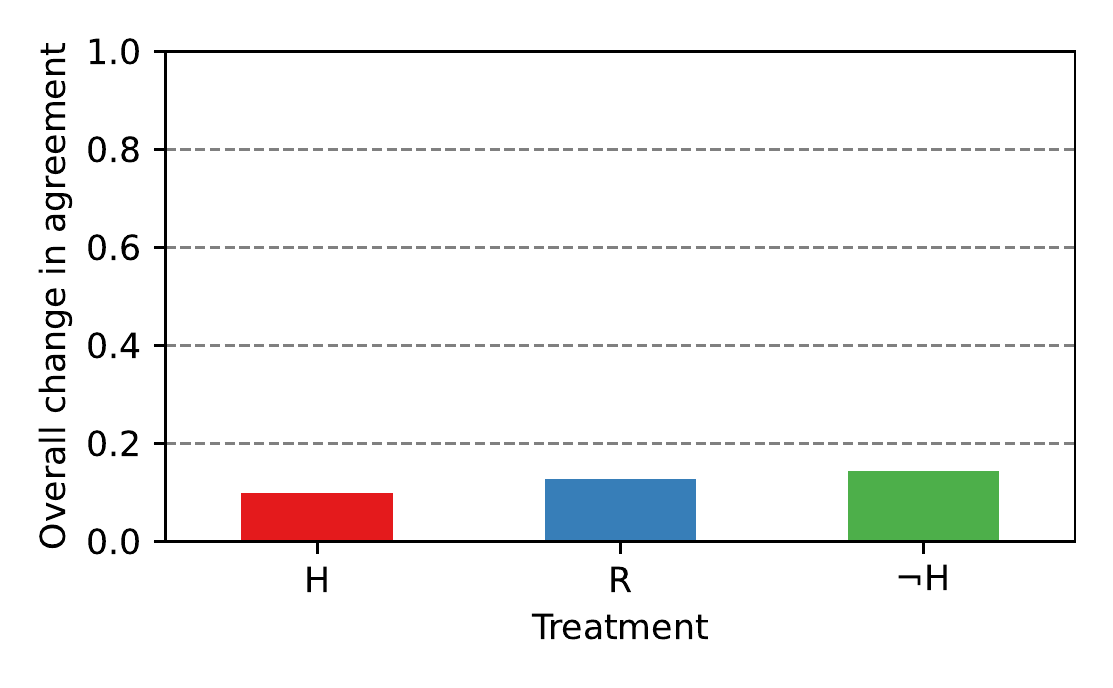}
        \caption{Overall Change in Agreement}
        \label{Fig:agreement_change}
    \end{subfigure}%

    \begin{subfigure}{.33\textwidth}
        \centering
        \includegraphics[width=0.99\textwidth]{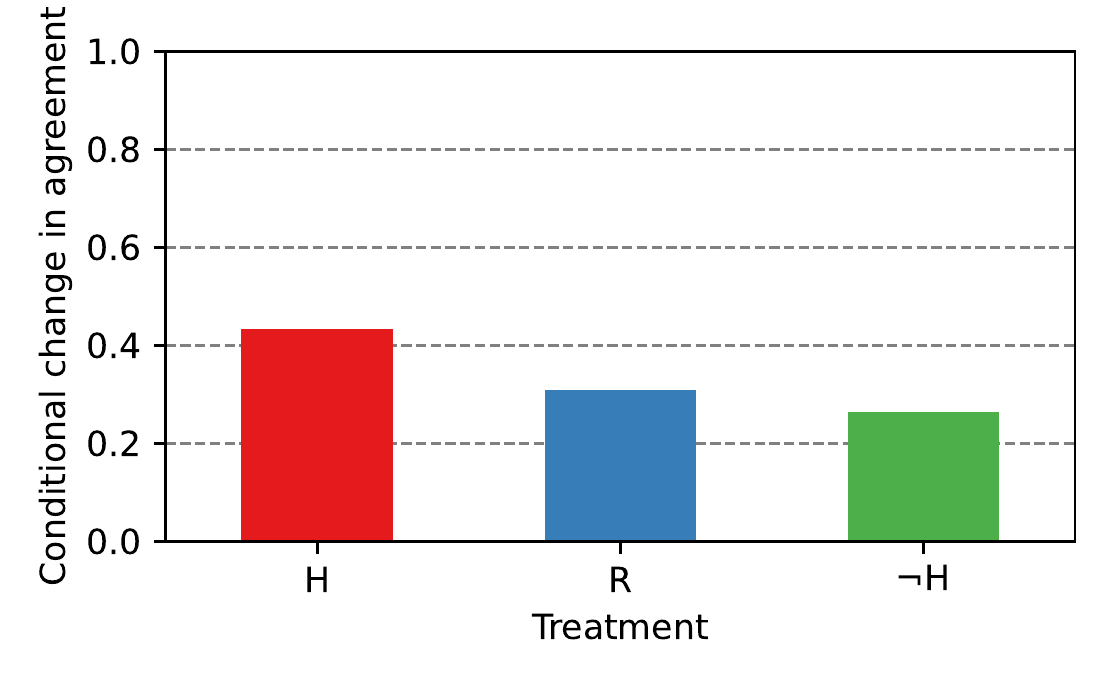}
        \caption{Conditional Change in Agreement}
        \label{Fig:agreement_change_conditional}
    \end{subfigure}%
    \begin{subfigure}{.33\textwidth}
        \centering
        \includegraphics[width=0.99\textwidth]{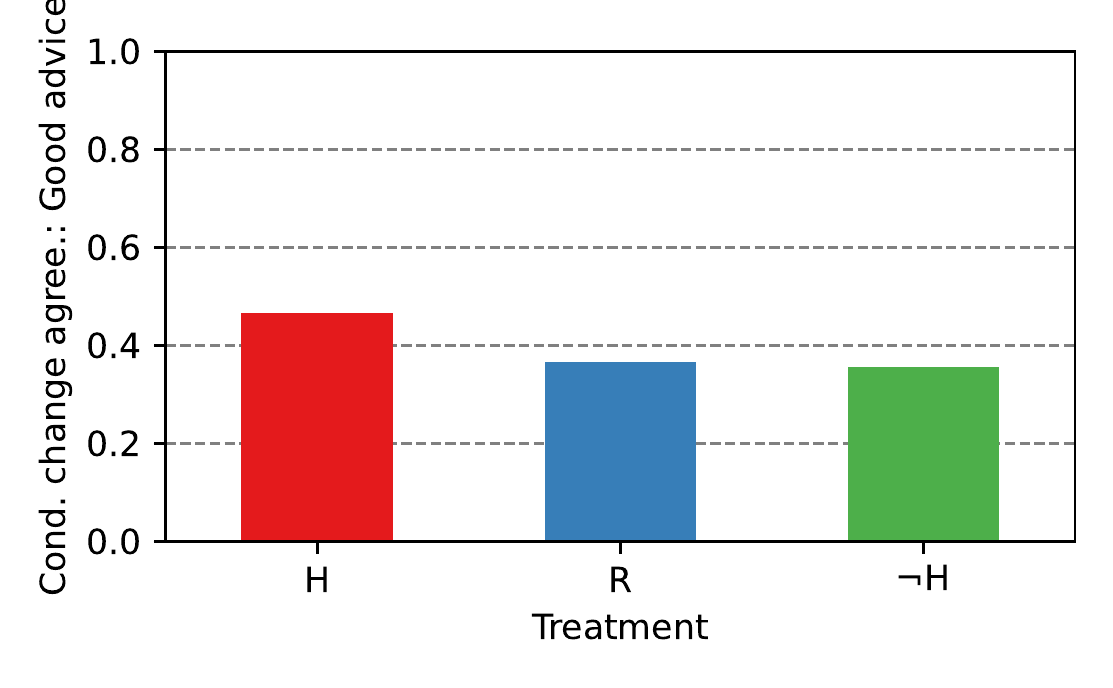}
        \caption{Conditional Change: Good Advice}
        \label{Fig:agreement_change_conditional_good}
    \end{subfigure}%
    \begin{subfigure}{.33\textwidth}
        \centering
        \includegraphics[width=0.99\textwidth]{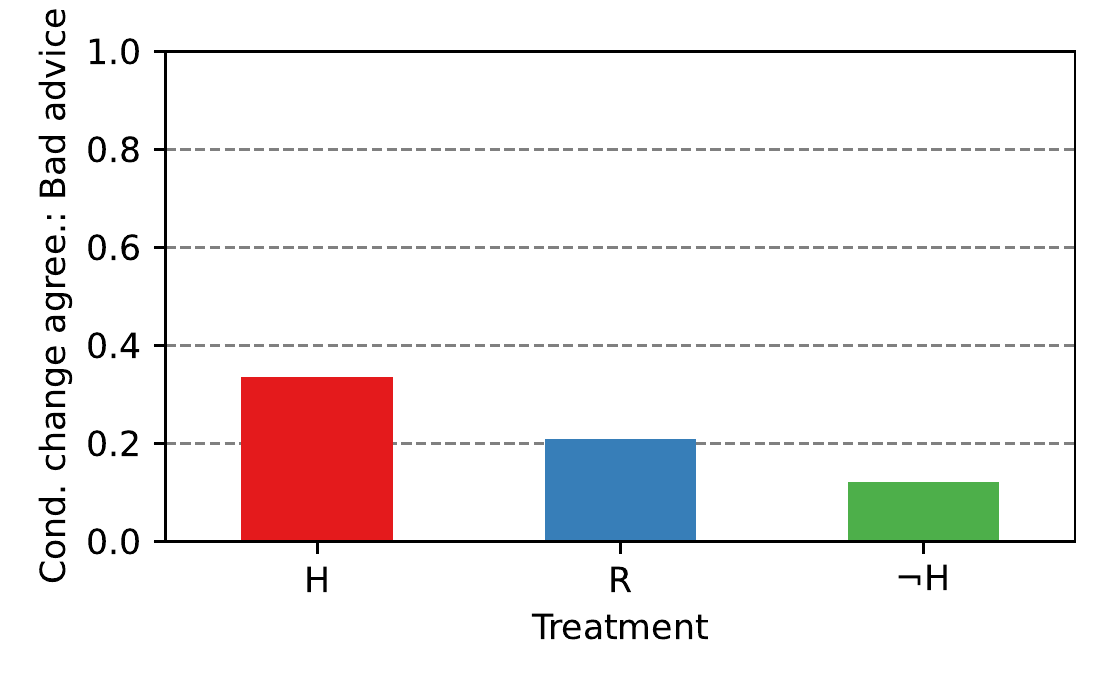}
        \caption{Conditional Change: Bad Advice}
        \label{Fig:agreement_change_conditional_bad}
    \end{subfigure}%
    \vspace{-2mm}
    \caption{Influence of machine advice on respondents' \textbf{agreement} with the advice, for decision aids with human-like (H), anti human-like ($\neg$H), and randomly distributed (R) errors. Pre and post-advice agreement, as well as the conditional change in agreement are higher for human-like decision aids than for anti human-like decision aids. On the other hand, for the overall change in agreement, the opposite pattern holds. Figures \ref{Fig:agreement_change_conditional_good} and \ref{Fig:agreement_change_conditional_bad} compare the conditional change in agreement for good and bad advice separately.}
    \label{Fig:agreement}
\end{figure*}

\subsection{Human Receptiveness to Machine Advice} \label{subsec:advice_taking}

After discussing the effect of human-machine similarity on human \emph{perceptions}, we move on to commenting on human advice taking \emph{behavior}. 
As a first sanity check, we test if machine advice had an effect in the expected direction. As shown in Table \ref{Tab:agreement_dist}, we found that many people updated their decisions after receiving opposing advice (the likelihood of doing so varies across treatments, as discussed below), but very few people changed their pre-advice decision after observing that the machine agrees with them. Namely, in the third row we see that very few respondents ($\leq1\%$) initially agreed with machine advice and then switched their decision and disagreed with it after observing confirming advice. I.e., machine advice affects people's decisions in the expected direction. 
Next, we explore the variation across treatments.

Figure \ref{Fig:agreement_pre} shows how the agreement of human \emph{pre-advice} decisions with machine advice varies across decision aids. As intended by the design of our decision aids, pre-advice agreement is positively correlated with human-machine similarity. Using a linear mixed model and a subsequent Wald test, we confirmed that this effect is statistically significant, with a p-value $<0.001$. This sanity check demonstrates that our decision aids successfully achieve their designated degree of similarity to human decisions.

After this sanity check, we move on to analyzing participants' \emph{post-advice} decisions. Figure \ref{Fig:agreement_post} show that people's post-advice decisions follow the same pattern as their pre-advice decisions. I.e., even after receiving machine advice, people's decisions are more similar to the human-like decision aids than to anti human-like decision aids. Using a linear mixed model and a subsequent Wald test, we found that this effect is also significant, with a p-value $<0.001$.

In Figure \ref{Fig:agreement_change}, we focus on the \emph{overall change in agreement}, calculated as the difference between post and pre-advice agreement. While both pre and post-advice decisions showed a positive correlation between human-machine similarity and agreement, the change in agreement exhibits the opposite pattern. Human-like decision aids lead to a slightly lower (by 5 percentage points) overall increase in agreement than anti human-like ones. This finding is corroborated by the regression and Wald test in Table \ref{Tab:agreement_change}, Model 1, with a p-value $<0.001$. This gives us

\begin{quote}
    \xhdr{Result 2} Anti human-like decision aids have a higher overall influence than human-like decision aids.
\end{quote}

However, it is important to keep in mind that the pre-advice agreement rates for anti human-like decision aids were significantly lower than for human-like ones. Hence, people had more opportunities to receive, and consequently take, opposing advice from anti human-like decision aids than from human-like ones. Next, we account for this.

In Figure \ref{Fig:agreement_change_conditional}, we calculate the \emph{conditional change in agreement} as the difference between post and pre-advice agreement, for instances where people initially disagreed with machine advice. I.e., this measure captures the amount of received opposing advice that was taken. With this, we control for the differences in pre-advice agreement rates across treatments. Even though the overall change in agreement was higher for anti human-like decision aids, the conditional change is higher for human-like decision aids. Specifically, the likelihood of taking opposing advice from human-like decision aids was 17 percentage points higher than for anti human-like ones (0.43$\%$ vs 0.26$\%$). This result is corroborated by the regression and Wald test shown in Table \ref{Tab:agreement_change_conditional}, Model 1, with a p-value $<0.001$. This leads to
    
\begin{quote}
    \xhdr{Result 3} Human-like decision aids have a higher conditional influence than anti human-like decision aids.
\end{quote}

\begin{figure*}[t]
    \centering
    \begin{subfigure}{.33\textwidth}
        \centering
        \includegraphics[width=0.99\textwidth]{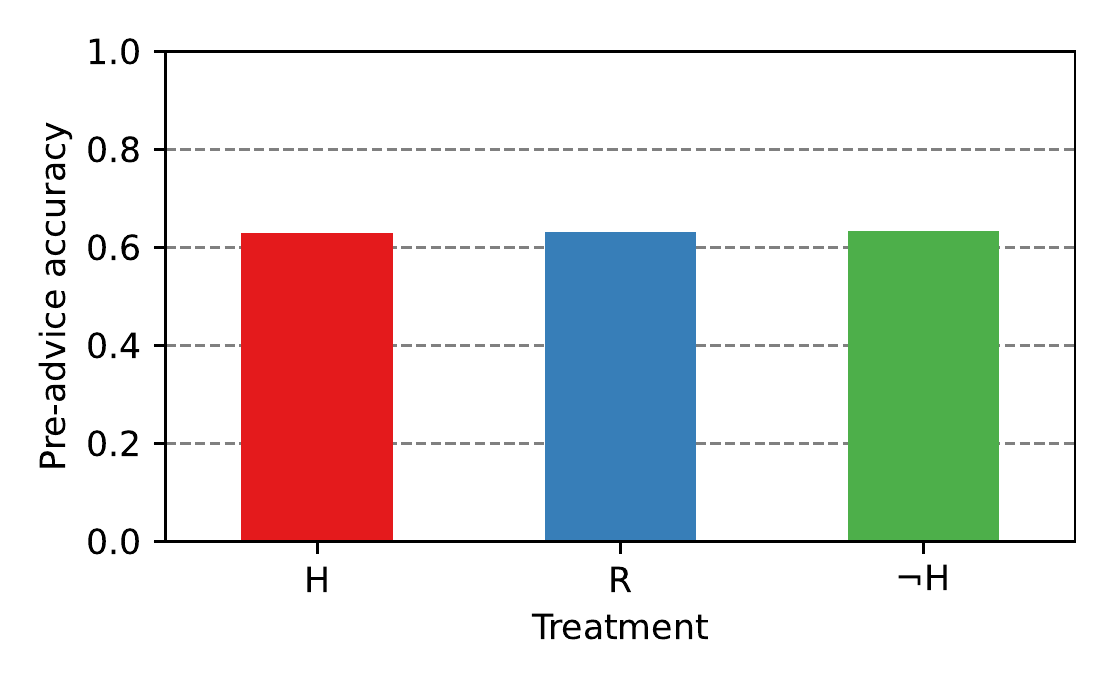}
        \caption{Pre-advice Accuracy}
        \label{Fig:accuracy_pre}
    \end{subfigure}%
    \begin{subfigure}{.33\textwidth}
        \centering
        \includegraphics[width=0.99\textwidth]{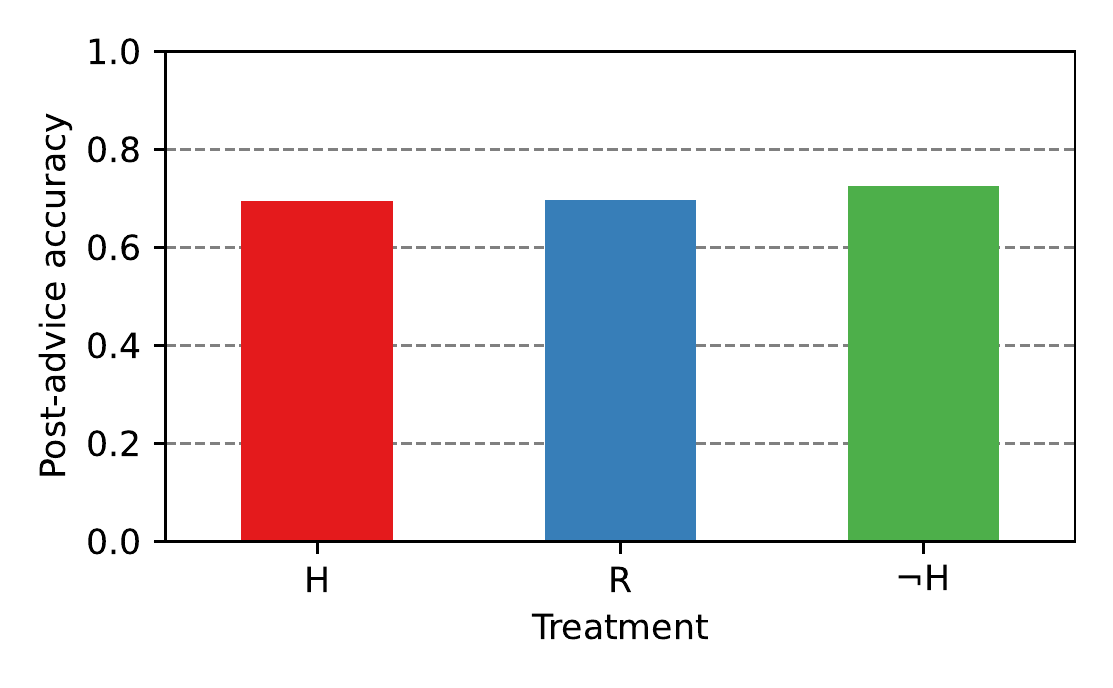}
        \caption{Post-advice Accuracy}
        \label{Fig:accuracy_post}
    \end{subfigure}%
    \begin{subfigure}{.33\textwidth}
        \centering
        \includegraphics[width=0.99\textwidth]{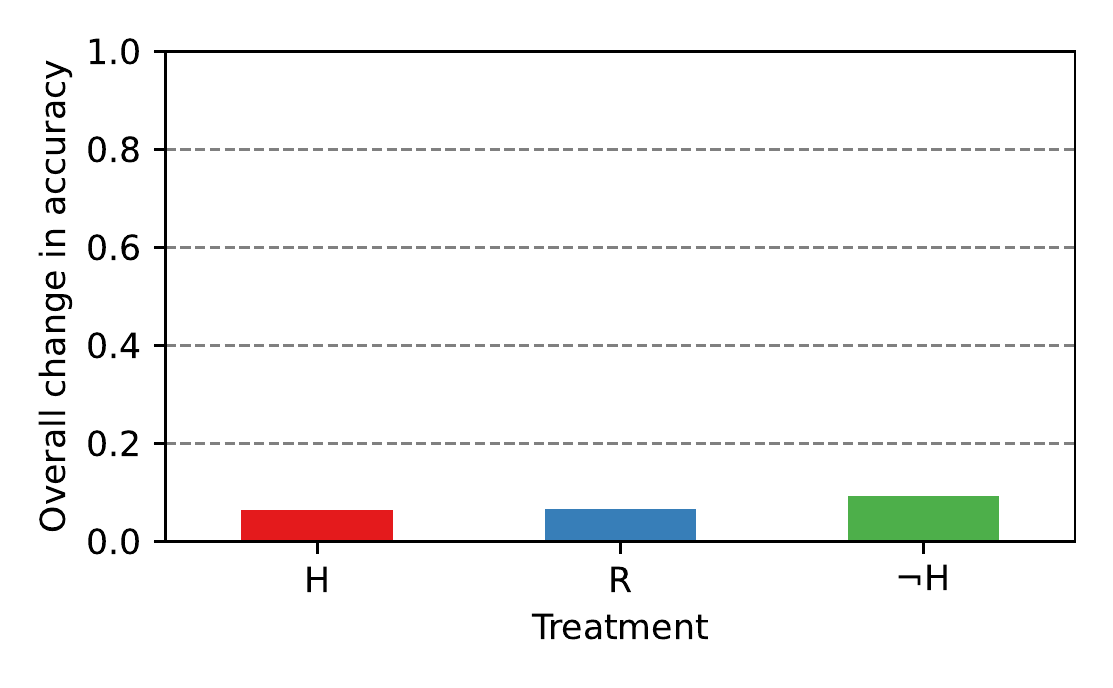}
        \caption{Overall Change in Accuracy}
        \label{Fig:accuracy_change}
    \end{subfigure}%

    \begin{subfigure}{.33\textwidth}
        \centering
        \includegraphics[width=0.99\textwidth]{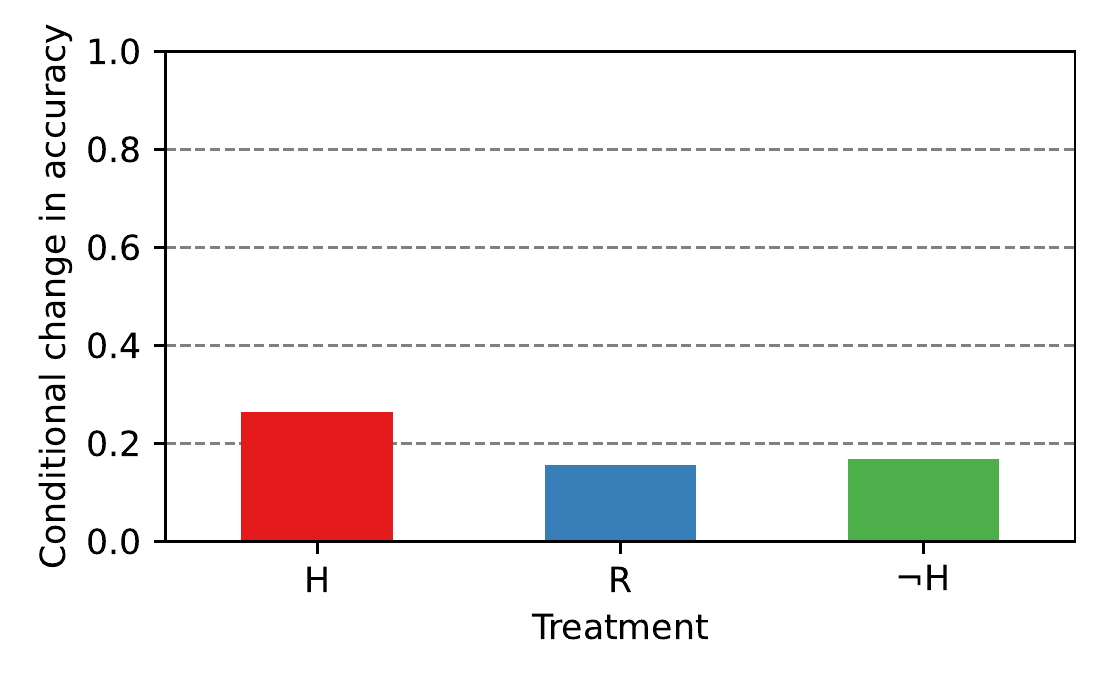}
        \caption{Conditional Change in Accuracy}
        \label{Fig:accuracy_change_conditional}
    \end{subfigure}%
    \hspace{25pt}
    \begin{subfigure}{.33\textwidth}
        \centering
        \includegraphics[width=0.99\textwidth]{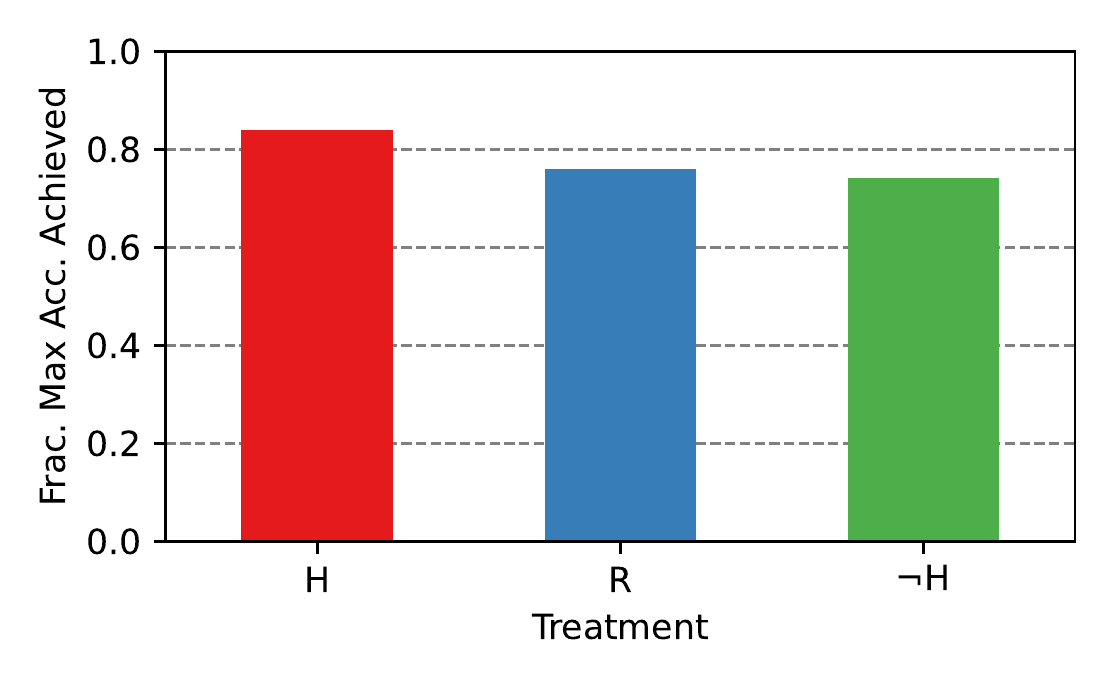}
        \caption{Fraction of Potential Achieved}
        \label{Fig:accuracy_frac_max}
    \end{subfigure}%
    \vspace{-2mm}
    \caption{Influence of machine advice on respondents' \textbf{accuracy}, for decision aids with human-like (H), anti human-like ($\neg$H), and randomly distributed (R) errors. Pre and post-advice accuracy, and the conditional change in accuracy are higher for human-like decision aids than for anti human-like decision aids. Human-like decision aids were also closer to reaching their full potential in terms of improving the respondents' accuracy. However, for the overall change in accuracy, the opposite holds. }
    \label{Fig:accuracy}
\end{figure*}

\xhdr{Control Variables}
All of the findings hold when we control for the decision-making task (i.e., for the dataset), the decision aid's accuracy, and the experimental phase. This can be seen in Figure \ref{Fig:agreement_separated} in the Appendix, where the results from Figure \ref{Fig:agreement} are separated by the three aforementioned control variables. We show that our results also remain statistically significant when we introduce these control variables in Tables \ref{Tab:agreement_change} and \ref{Tab:agreement_change_conditional}, Model 2.

As was the case for perceptions, the control variables again have a significant effect on advice taking behavior. As seen in Tables \ref{Tab:agreement_change} and \ref{Tab:agreement_change_conditional}, Model 2, the control variables have a consistent effect on both overall and conditional agreement. The coefficients indicate that people were more likely to take machine advice in the second experimental phase.\footnote{This difference could be caused by the introduction of monetary incentives and removal of feedback about performance in the second phase, or other factors, such as learning or fatigue effects.} Also, people were more likely to take advice from more accurate decision aids. There was no significant difference between advice taking for the COMPAS and Speed Date dataset, for which people have a similar baseline accuracy. However, for the Age dataset, where their baseline accuracy was higher, people were less likely to take machine advice.
% !TEX root = error_types.tex

\begin{table}[t]
    \centering
    \small
    {\def\sym#1{\ifmmode^{#1}\else\(^{#1}\)\fi}
    \begin{tabulary}{\columnwidth}{L|RR}
    \toprule
    & \multicolumn{2}{c}{\textbf{Model}} \\
    \midrule
    Human-like Treatment & 0.127\sym{**} & (0.042)\\
    $\neg$ Human-like Treatment & -0.0883\sym{**} & (0.028)\\
    Advice Correct & 0.156\sym{***}& (0.028)\\
    Human-like \# Advice Correct & -0.0265 & (0.041)\\
    $\neg$ Human-like \# Advice Correct & 0.0788\sym{**} & (0.030)\\
    Intercept & 0.210\sym{***}& (0.025)\\
    \bottomrule
    \end{tabulary}}
    \caption{Dependent variable: Conditional agreement change. This Model builds up on Model 1 from Table \ref{Tab:agreement_change_conditional} by accounting for the correctness of machine advice. Interactions are denoted by \#. The reference category is the Random treatment. Standard errors in parentheses. * symbols next to coefficients indicate their statistical significance as follows: * $p<0.05$, ** $p<0.01$, *** $p<0.001$. N = 12578.}
    \label{Tab:agreement_change_good_vs_bad}
\end{table}

\begin{figure*}[t]
    \centering
    \begin{subfigure}{.33\textwidth}
        \centering
        \includegraphics[width=0.99\textwidth]{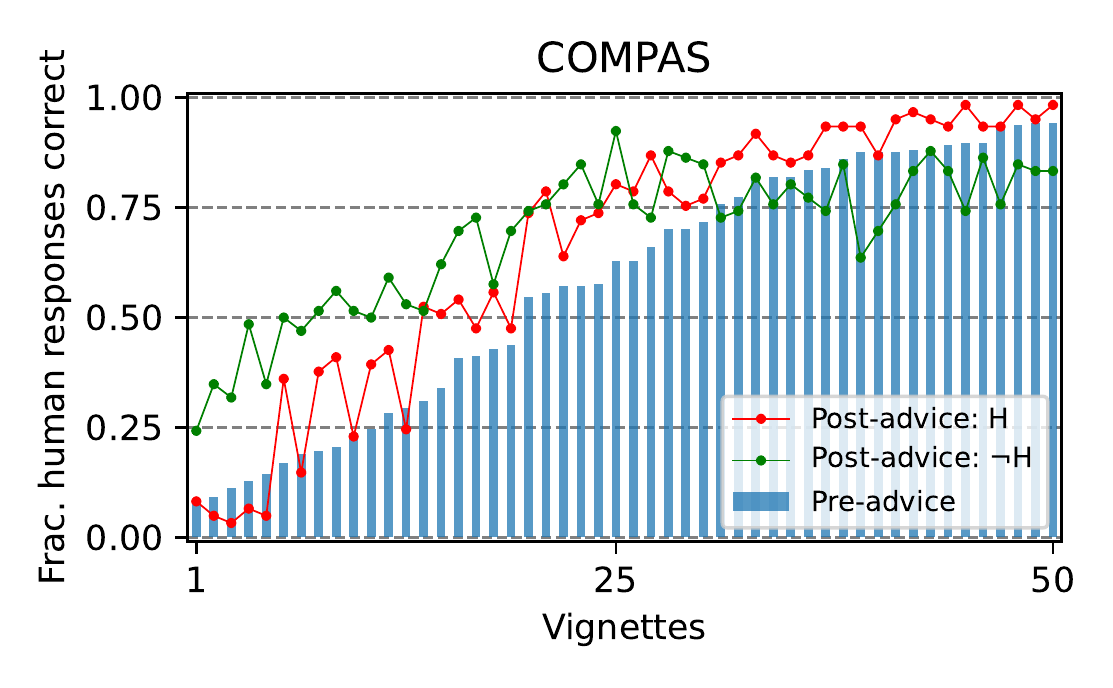}
        \caption{COMPAS}
        \label{Fig:accuracy_per_vignette_COMPAS}
    \end{subfigure}%
    \begin{subfigure}{.33\textwidth}
        \centering
        \includegraphics[width=0.99\textwidth]{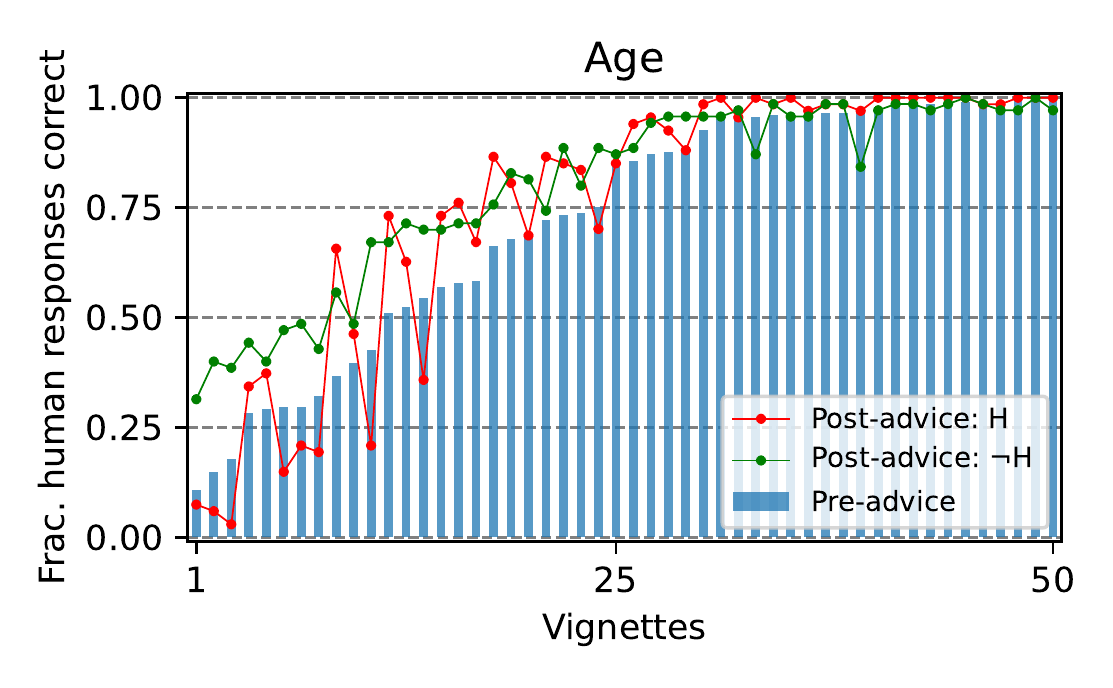}
        \caption{Age}
        \label{Fig:accuracy_per_vignette_age}
    \end{subfigure}%
    \begin{subfigure}{.33\textwidth}
        \centering
        \includegraphics[width=0.99\textwidth]{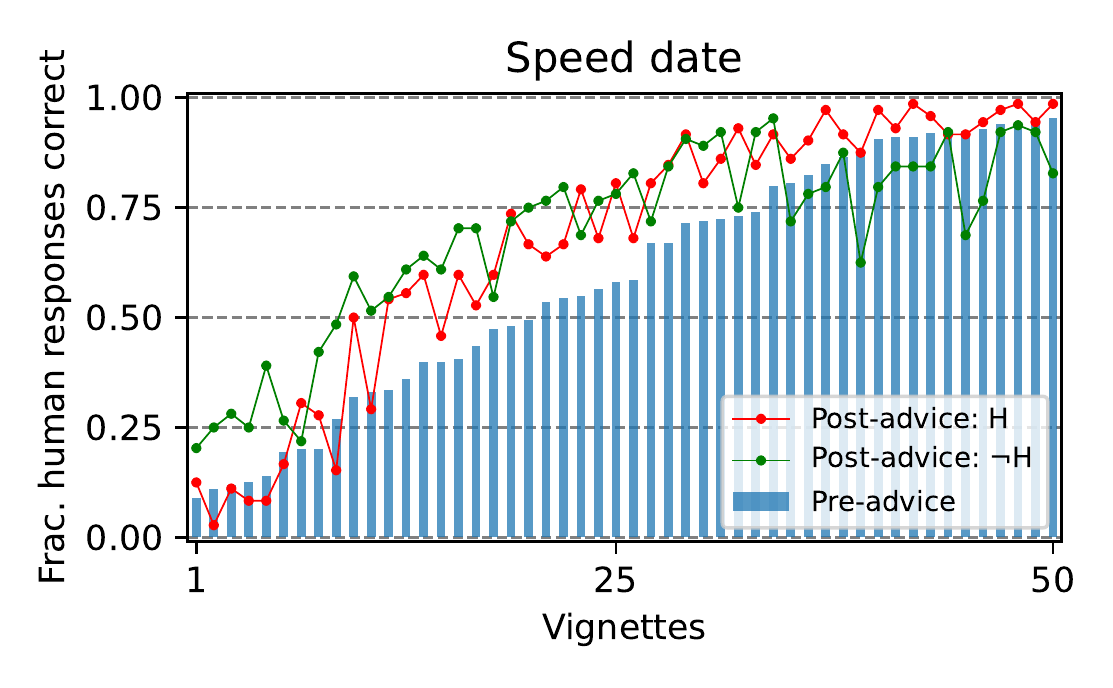}
        \caption{Speed Date}
        \label{Fig:accuracy_per_vignette_speed_date}
    \end{subfigure}%
    \caption{Influence of machine advice on respondents' accuracy. The bars show the respondents' pre-advice accuracy, while the lines show their post-advice accuracy, for decision aids with human-like (red line) and anti human-like (green line) errors. On the x-axis, the 50 vignettes are sorted increasingly w.r.t. the fraction of respondents' whose pre-advice decision was correct. We observe that on the left side of the plots (where people are more prone to errors) the anti human-like decision aid impacts the respondents' accuracy more positively on average. On the right side of the plots (where people are less likely to make errors) the human-like decision aid improves accuracy, while the anti human-like one reduces people's accuracy.}
    \label{Fig:accuracy_per_vignette}
\end{figure*}

\subsection{Impact of Advice on Human Accuracy} \label{subsec:accuracy}

In this Section, we conduct an exploratory study, and describe the impact of human-machine similarity on the accuracy of people's decision in our experiments. 

\xhdr{Comparing the Influence of Good and Bad Advice}
The impact of people's advice taking behavior on the accuracy of their decisions depends not only on the amount of advice that is taken, but also on the quality of the taken advice. In Figures \ref{Fig:agreement_change_conditional_good} and \ref{Fig:agreement_change_conditional_bad}, we show the conditional change in agreement for correct and incorrect advice separately. Specifically, Figure \ref{Fig:agreement_change_conditional_good} depicts the influence of good advice, i.e., the fraction of times people change their incorrect pre-advice decision after observing correct machine advice. On the other hand, Figure \ref{Fig:agreement_change_conditional_bad} shows how people react to bad advice, i.e., the fraction of times people update their correct pre-advice decision after receiving incorrect advice.

Firstly, and reassuringly, we observe that good advice was more influential than bad advice, for all of the decision aids. However, we see differences in the magnitude of influence across treatments. Human-like decision aids have a higher conditional influence (i.e., people exhibit a higher propensity to take opposing advice), both for correct advice and for incorrect advice. I.e., people are more likely to take good advice from human-like decision aids, but they are more likely to reject bad advice from anti human-like decision aids. These observations are corroborated by the regression in Table \ref{Tab:agreement_change_good_vs_bad}. Hence, our results do not show evidence that varying the degree of human-machine similarity might help people distinguish between good and bad machine advice.

\xhdr{Impact on Accuracy}
Figure \ref{Fig:accuracy_pre} shows that, descriptively, people's \emph{pre-advice} accuracy is on average equal across treatments. I.e., as expected, before observing machine advice, people were equally accurate in all of our treatments. Using a linear mixed model, we found that the difference across treatments is indeed not significant, with a Wald test p-value of 0.7982.
However, the \emph{post-advice} accuracy slightly differs across treatments, as visible in Figure \ref{Fig:accuracy_post}. Using a linear mixed model, we found that post-advice accuracy is higher for the anti human-like treatment, with a borderline significant Wald test p-value $<0.05$.

In Figure \ref{Fig:accuracy_change}, we focus on the \emph{overall change in accuracy}, calculated as the difference between post and pre-advice accuracy. Consistently with our finding about the post-advice accuracy, we observe that the anti human-like treatment leads to a slightly higher overall change in accuracy. Using the same statistical analysis as above, we found that this effect is also borderline significant, with a p-value $<0.05$.

Next, we discuss the \emph{conditional change in accuracy}, calculated as the difference between post and pre-advice accuracy, for instances where a respondent's pre-advice decision disagreed with machine advice. In Figure \ref{Fig:accuracy_change_conditional}, we observe that human-like decision aids lead to a higher conditional change in accuracy. Again, this effect is found to be borderline statistically significant, with a p-value $<0.05$.

All of the findings related to people's accuracy are in line with the findings on people's advice taking behavior, from Section \ref{subsec:advice_taking}. Since the decision aids are on average more accurate than humans (i.e., they provide more good advice than bad advice), and good advice was more influential than bad advice, machine advice positively impacted the accuracy of people's responses in line with their advice taking propensity. The overall influence of anti human-like advice is higher, and so is the overall change in accuracy. On the other hand, the conditional influence of human-like advice is higher, and so is the conditional change in accuracy.

While Figure \ref{Fig:accuracy} shows the decision aids' impact on accuracy averaged across all vignettes, in Figure \ref{Fig:accuracy_per_vignette} we show how the decision aids impacted people's accuracy for specific vignettes (averaged across the two experimental phases and degrees of accuracy). We observe that for vignettes where most people made errors (left side of the plots), the human-like decision aid (red line) also made errors and hence the anti human-like (green) decision aid had a more positive impact on the respondents' accuracy. However, for instances where most people made correct predictions (right side of plots), the human-like (green) decision aids helped many of the remaining respondents correct their decisions, unlike the anti human-like (green) decision aid which steered some of the respondents away from their initially correct responses.

\xhdr{Potential for Improving Accuracy}
Next, we go back to the illustration from Figure \ref{Fig:similarity_vs_complementarity}. Using the same reasoning as the one applied in the illustration, we can conclude that anti human-like decision aids have the potential to improve the accuracy of human decisions more than human-like decision aids. I.e., for cases where humans make incorrect predictions, anti human-like decision aids are more likely to be able to give people correct advice. In the following analysis, we try to account for this. We explore how close the post-advice accuracy of human decisions is to the upper bound given by the union of the responses where humans were correct and those where the decision aids were correct. 

Figure \ref{Fig:accuracy_frac_max} shows the results of this analysis. This plot shows the fraction of correct post-advice decisions, amongst all of the decisions where either the decision aid's advice was correct or the human respondent's pre-advice decision was correct. This analysis aims to capture the fraction of the upper bound on accuracy that was achieved. Here we found that human-like decision aids reached more of their full potential than anti human-like decision aids. This difference was confirmed to be statistically significant, with a p-value $<0.001$. I.e., even though the overall increase in accuracy was slightly higher for anti human-like decision aids (Fig. \ref{Fig:accuracy_change}), they were significantly farther from reaching their full potential for improving human decisions.

% !TEX root = error_types.tex

\section{Discussion} \label{sec:discussion}

\xhdr{Limitations and Future Work}
In our experiments, we considered synthetic decision aids with three degrees of similarity to human decisions, three binary prediction tasks, and two degrees of predictive accuracy. Future work could study additional variables or more fine-grained levels of these variables. E.g., the human-like decision aids we studied made mistakes only where humans made them as well, while the anti human-like decision aids made mistakes only where most humans were accurate. These differences may be less severe for real-world decision aids, and it is hence worth conducting further research varying the degree of human-machine similarity beyond the three degrees we studied.

We compared decision aids that make mistakes for different inputs. In a multi-class (e.g., emotion recognition) or regression (e.g. real-estate price or continuous age estimation), setting, one could also compare decision aids that make different types of mistakes for the same input. E.g., people and human-like decision aids might underestimate the true value of a certain vignette in a regression, while anti human-like decision aids might overestimate it.

Future studies could also consider other respondent samples, e.g., representative of the US population or other non-US populations. For decision aid used by professionals in the real world (e.g., the COMPAS tool for predicting criminal recidivism used by judges in the US legal system), it may also be interesting to test whether our findings hold for the relevant population of real world decision makers.

In our work we focused on experimentally testing the effect of human-machine similarity on human perceptions and utilization of machine decision aids. Past research has identified other factors that also influence how people take machine advice, such as the interpretability and explainability of machine advice \cite{poursabzi2018manipulating, wang2021explanations, zhang2020effect}, and future research could explore if these factors moderate the effects of human-machine similarity. Finally, future work in social psychology could study the psychological mechanisms which underlie the observed effect of human-machine similarity on advice taking. As a first step in this direction, we provide a brief discussion on this below.

\xhdr{Underlying Psychological Mechanisms}
As reported in Section \ref{subsec:perceptions}, people perceive human-like decision aids as more predictable, accurate and useful. The predictability of others' future actions has long been recognized as a crucial component of interpersonal trust \cite{rempel1985trust}, and was also shown to be critical for human trust in automation \cite{hoff2015trust, madhavan2007similarities}. People's receptiveness to advice was also shown to be correlated with the inferred quality and accuracy of an advisor's advice \cite{bonaccio2006advice}. Finally, the perceived usefulness of information technologies is highly predictive of the adoption of such technologies \cite{davis1989perceived, karahanna1999psychological}. Hence, people's perceptions about the comparative advantages of human-like decision aids in terms of their predictability, accuracy and usefulness may lead to the observed higher receptiveness to opposing advice given by such systems.

One crucial mediator of the effect of human-machine similarity on advice taking could be trust, which is found to positively affect receptiveness to advice \cite{bonaccio2006advice}. As mentioned in Section \ref{sec:related_work}, literature on algorithmic aversion has found that people tend to lose trust in algorithms more quickly than in human advisors after observing them make the same mistakes \cite{dietvorst2015algorithm}. We comment on two concepts which are found to affect trust resilience, which go beyond the human/algorithmic identity of the advisor and may explain our observations: error severity and anthropomorphism. 

The magnitude of a system's errors was shown to be correlated with the magnitude of users' loss of trust \cite{rossi2017timing, weun2004impact}. It is possible that human-like mistakes are perceived as less severe, while anti human-like errors are perceived as egregious, hence leading to lower trust resilience for complementary decision aids.
Anthropomorphism refers to the tendency to ascribe human-like characteristics to non-human agents \cite{epley2007seeing}. Prior research has found that a robot's behavior affects the degree of anthropomorphism \cite{duffy2003anthropomorphism, zlotowski2015anthropomorphism}. More anthropomorphic machine advisors were observed to exhibit higher degrees of trust resilience \cite{de2016almost}. Hence, decision aids which make mistakes more similar to human ones may also be anthropomorphized more, and in turn be trusted more.

\xhdr{Perceptions vs Behavior}
Next we briefly comment on the relationship between people's perceptions about the decision aids (covered in Section \ref{subsec:perceptions}) and their advice taking behavior (discussed in Sections \ref{subsec:advice_taking} and \ref{subsec:accuracy}).

People perceived human-like decision aids as more \emph{useful} than anti human-like decision aids. To explore if people's perceptions of usefulness correspond to the observed usefulness of the decision aids in practice, it is necessary to define what constitutes usefulness in this setting. One possible definition could be that machine advice is useful if it leads to an increase in the accuracy of people's decisions.\footnote{Other possible definitions of usefulness might aim to capture the ease and speed of decision-making, or other similar factors.} In our experiments, human-like decision aids --- which were perceived as more useful --- were more successful in reaching their potential for improving accuracy. However, the overall increase in accuracy was slightly larger for anti human-like decision aids, and hence perceptions about the decision aids' usefulness arguably did not coincide with the observed impact of the advice on the accuracy of people's decisions.

Human-like decision aids were also perceived as more \emph{accurate} than anti human-like decision aids. Since all of the compared decision aids were equally accurate, people's perceptions did not coincide with the factual reality.

Finally, in terms of the perceived \emph{predictability} of machine advice, human-like decision aids again received higher ratings. Given the higher overlap between people's decisions and those of human-like decision aids, it seems plausible that people were able to predict human-like machine advice better. In future research, it would be interesting to test whether people are actually better in predicting such advice.

\xhdr{Design Implications}
With the increasing popularity of ML algorithms that aim to complement human skills, it is important to understand the effects of human-machine complementarity on machine-assisted decision-making. Our findings about the effects of human-machine similarity on people's perceptions and utilization of machine advice have important implications for the design of decision aids, particularly in settings where human agents are the final decision makers, while algorithmic decision aids serve as advisors.

Depending on the normative goals of utilizing machine assistance, it may be beneficial to use decision aids with different degrees of human-machine similarity. To ensure that a decision aid has a high influence for specific pieces of advice (e.g., a set of especially important or sensitive decisions), one may opt for human-like decision aids. The same holds if the normative goal is to ensure that people perceive the decision aid more favorably in terms of its usefulness, accuracy and predictability. On the other hand, if the goal is to maximize the overall influence of machine advice, a decision aid complementary to humans might be a better choice.

To tailor the degree of human-machine similarity to the normative goals of interest it is necessary to have access to models of human decision-making. Such models can be trained using data about people's past decisions in the relevant decision context. With access to models of human decision-making, developers can control the similarity between people's and algorithmic mistakes in various ways. One simple approach would entail leveraging predictive multiplicity \cite{marx2020predictive}. Namely, when multiple competing decision aids exhibit similar degrees of accuracy, one could select the decision aid based on the degree of human-machine similarity. Alternatively, when only one decision aid is available, one could manipulate the perceived similarity of human and algorithmic errors by selectively choosing when the algorithm provides advice (e.g., by avoiding to give advice for inputs where most people are predicted to (dis)agree with the machine advice). Finally, future research may enable the development of algorithms which in addition to optimizing for predictive accuracy can also optimize for the degree of similarity of human and machine decisions. This line of research fits well within existing efforts on developing machine learning algorithms which account for the presence of human agents in their learning procedure \cite{de2020regression, madras2018predict, meresht2020learning}.

\xhdr{Conclusion}
Our work contributes to the growing body of research on machine-assisted decision-making by studying a specific factor that may influence people's advice taking behavior: the degree of similarity between the decision aid's errors and typical human errors. In a series of large-scale online experiments, we experimentally show that human perceptions and utilization of algorithmic advice are in fact influenced by the similarity of human and machine errors. We invite future interdisciplinary research in social psychology and computer science, that will both provide deeper insights about the psychological mechanisms underlying our findings, and promote the development of algorithmic decision aids which may account for human-machine similarity.

\pagebreak
\section*{Ethics Statement}
The design of our human-subject experiments has been approved by our institution’s Ethical Review Board. The data was collected and stored in line with relevant ethical guidelines, which promote respondents' dignity, autonomy, and privacy (including but not limited to: paying respondents a fair wage, gathering respondents' informed consent, and avoiding storing personally identifiable data).

The purpose of our research is to identify factors which influence how people perceive and utilize machine decision aids, in order to support the design of decision aids that will effectively assist human decision-making. However, it is important to note that, as is the case for much research on human perceptions and behavior, these findings could also be utilized with malicious intent to steer people towards perceiving and utilizing decision aids in undesirable ways (e.g., over-relying on low quality decision aids, or under-relying on high quality decision aids).

\section*{Acknowledgements}
This research was supported in part by a European Research Council (ERC) Advanced Grant for the project "Foundations for Fair Social Computing", funded under the European Union’s Horizon 2020 Framework Programme (grant agreement no. 789373).

\bibliography{error_types}

\clearpage
\newpage
% !TEX root = error_types.tex

\section{Appendix} \label{sec:appendix}

In the following subsections we provide additional details about our experiments (Section \ref{subsec:additional_methodology}) and their results (Section \ref{subsec:additional_results}).

\subsection{Additional Details about the Experiments} \label{subsec:additional_methodology}

\begin{figure*}[t]
    \centering
    \begin{subfigure}{.49\textwidth}
        \centering
        \includegraphics[width=0.99\textwidth]{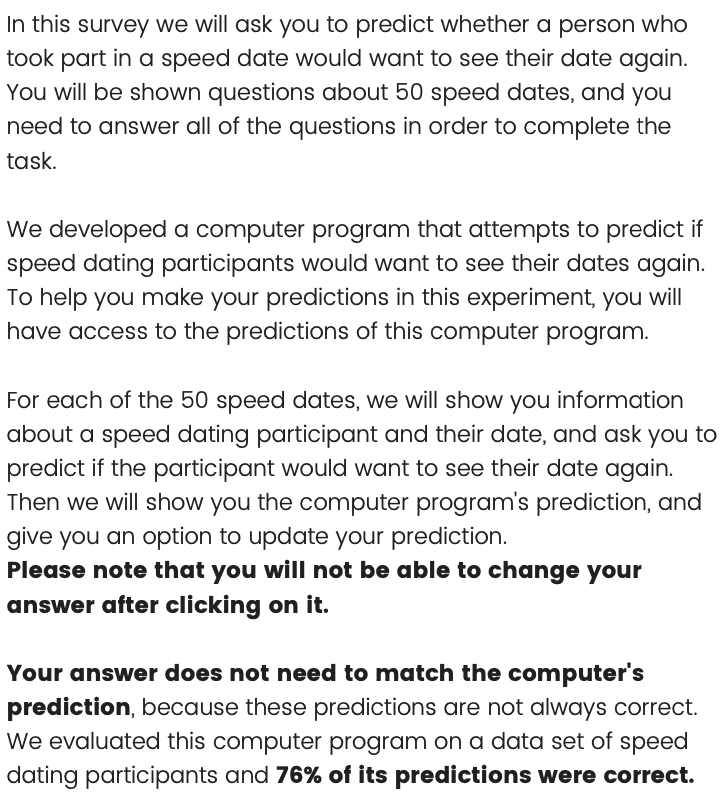}
        \caption{Description of the decision-making task}
        \label{Fig:intro_task}
    \end{subfigure}%
    \hfill
    \begin{subfigure}{.49\textwidth}
        \centering
        \includegraphics[width=0.99\textwidth]{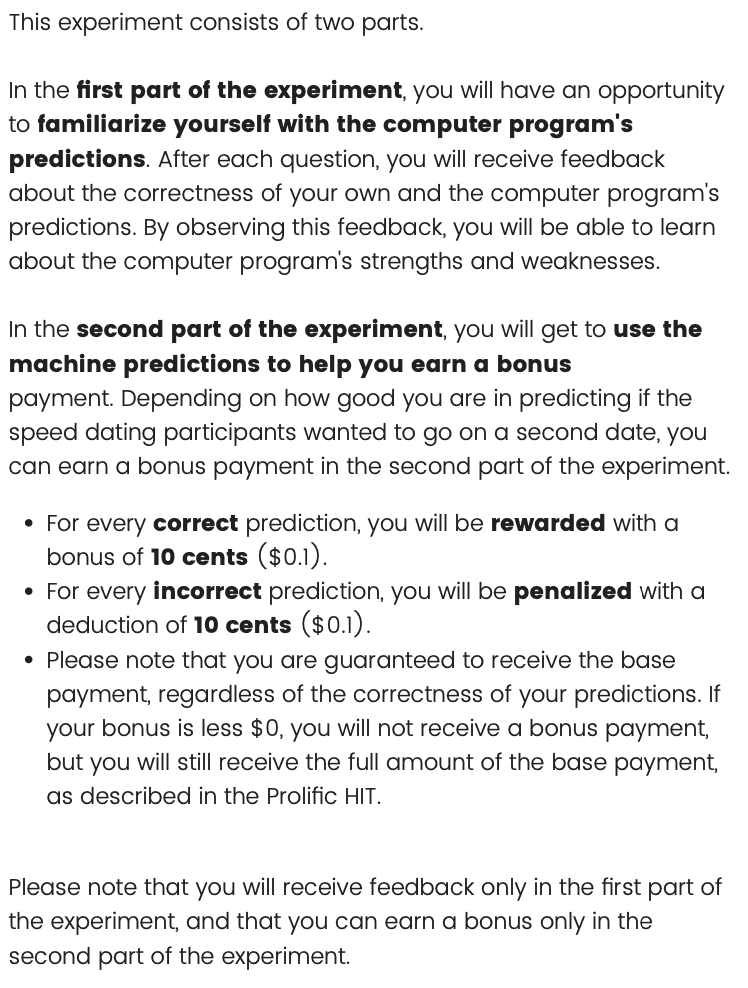}
        \caption{Description of the experimental setup, including monetary incentives}
        \label{Fig:intro_experiment}
    \end{subfigure}%
    \caption{Introductory text, shown to participants when they accessed the online survey}
    \label{Fig:survey_intro}
\end{figure*}

\begin{figure*}[t]
    \centering
    \begin{subfigure}{.49\textwidth}
        \centering
        \includegraphics[width=0.99\textwidth]{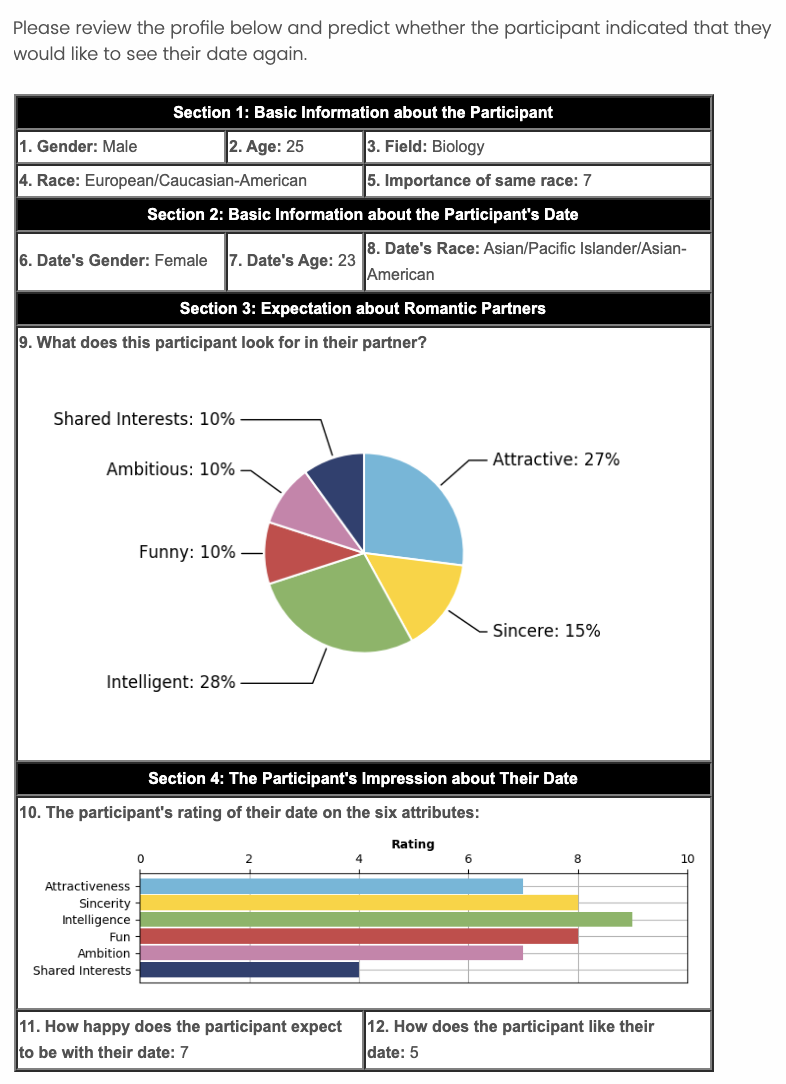}
        \caption{Vignette - Speed Date}
        \label{Fig:date_description}
    \end{subfigure}%
    \hfill
    \begin{subfigure}{.49\textwidth}
        \centering
        \includegraphics[width=0.99\textwidth]{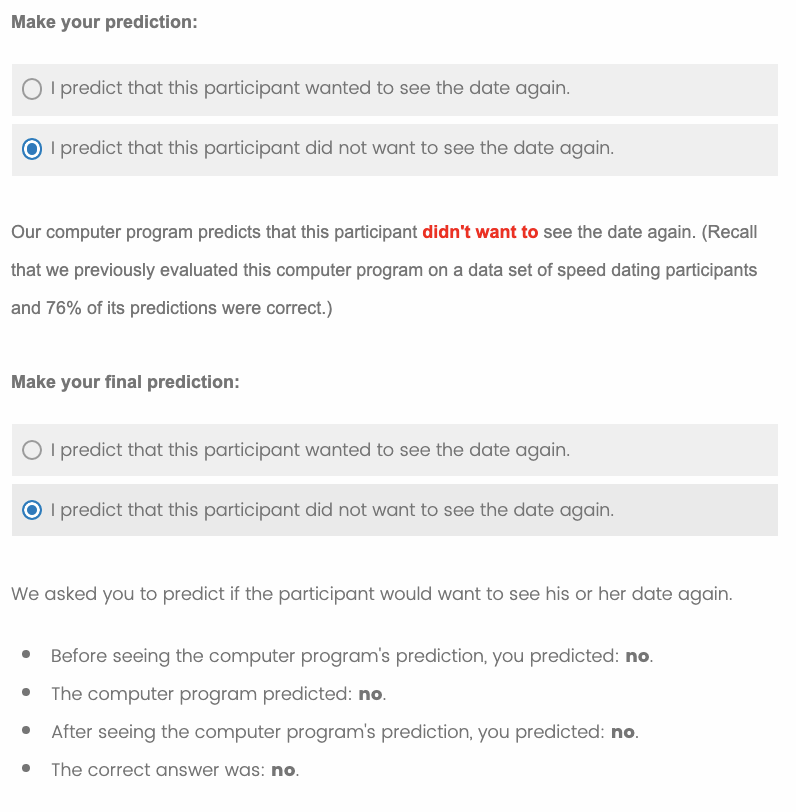}
        \caption{Pre-advice response options, followed by machine advice, and post-advice response options. The final part shows the feedback that was provided to participants after every question in the first phase of the survey.}
        \label{Fig:date_questions}
    \end{subfigure}%
    \\
    \begin{subfigure}{.49\textwidth}
        \centering
        \includegraphics[width=0.99\textwidth]{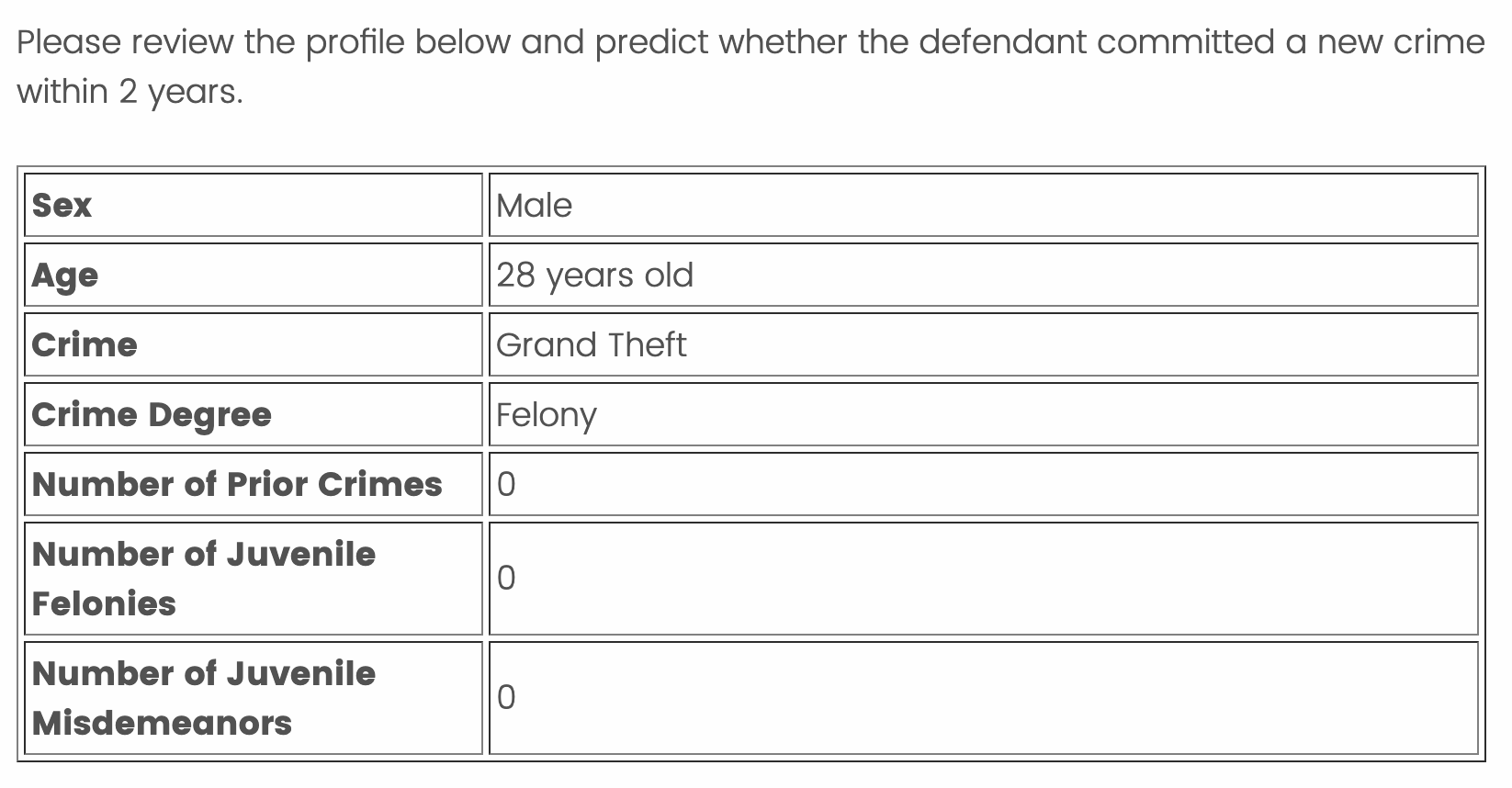}
        \caption{Vignette - COMPAS}
        \label{Fig:defendant_description}
    \end{subfigure}%
    \hfill
    \begin{subfigure}{.49\textwidth}
        \centering
        \includegraphics[width=0.99\textwidth]{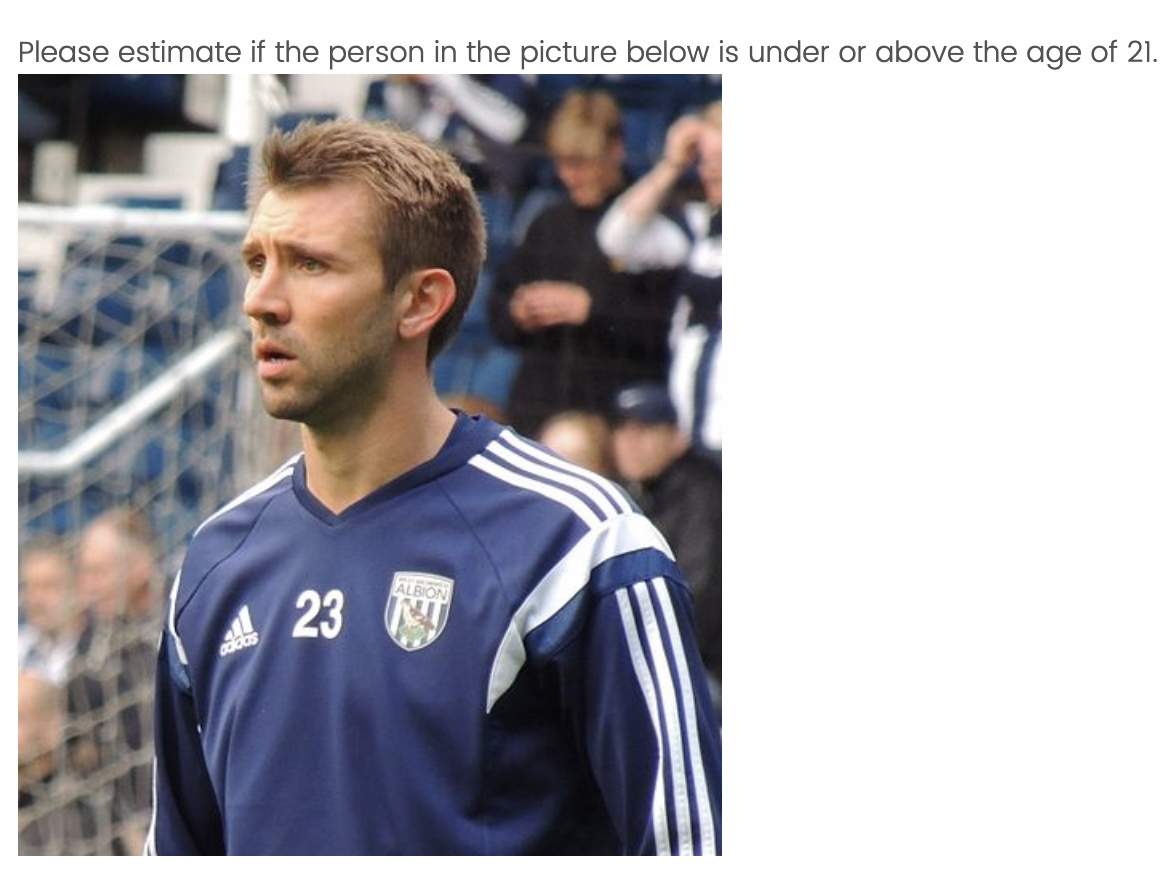}
        \caption{Vignette - Age}
        \label{Fig:age_picture}
    \end{subfigure}%
    \caption{Vignettes and response options}
    \label{Fig:survey_questions}
\end{figure*}

\begin{figure*}[t]
    \centering
    \begin{subfigure}{.49\textwidth}
        \centering
        \includegraphics[width=0.99\textwidth]{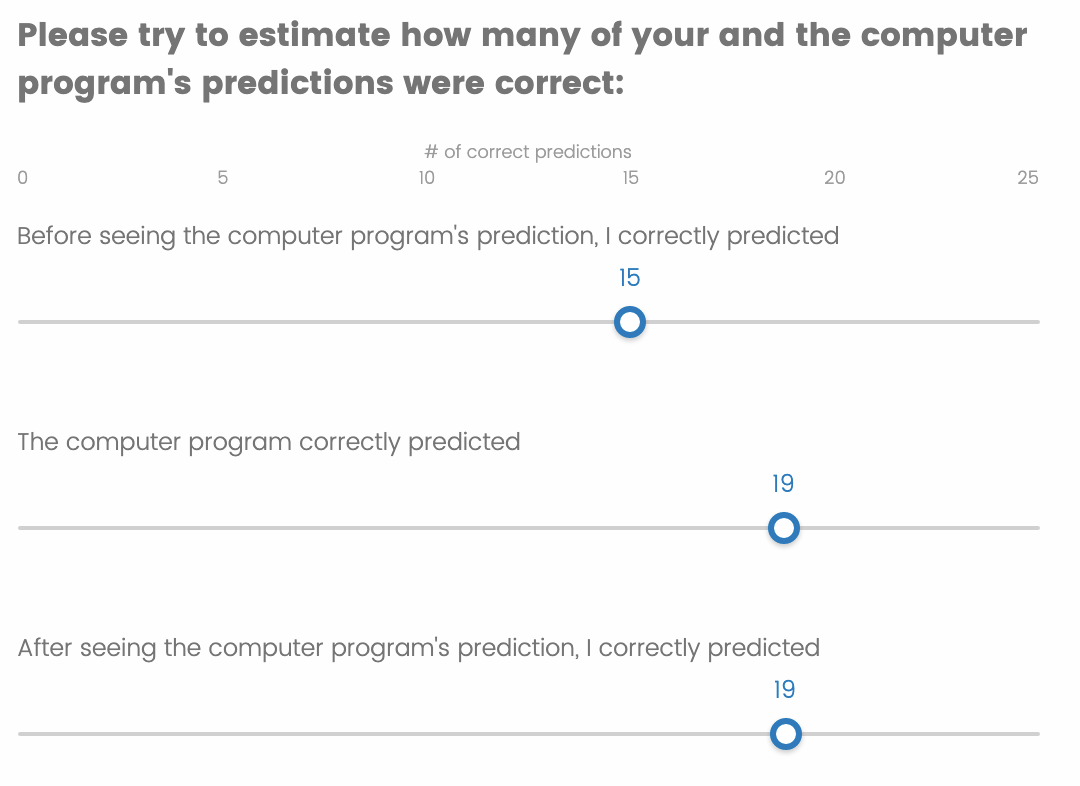}
        \caption{Estimating the accuracy of the machine predictions, as well as the respondent's personal pre and post-advice accuracy}
        \label{Fig:estimate_correctness}
    \end{subfigure}%
    \hfill
    \begin{subfigure}{.49\textwidth}
        \centering
        \includegraphics[width=0.99\textwidth]{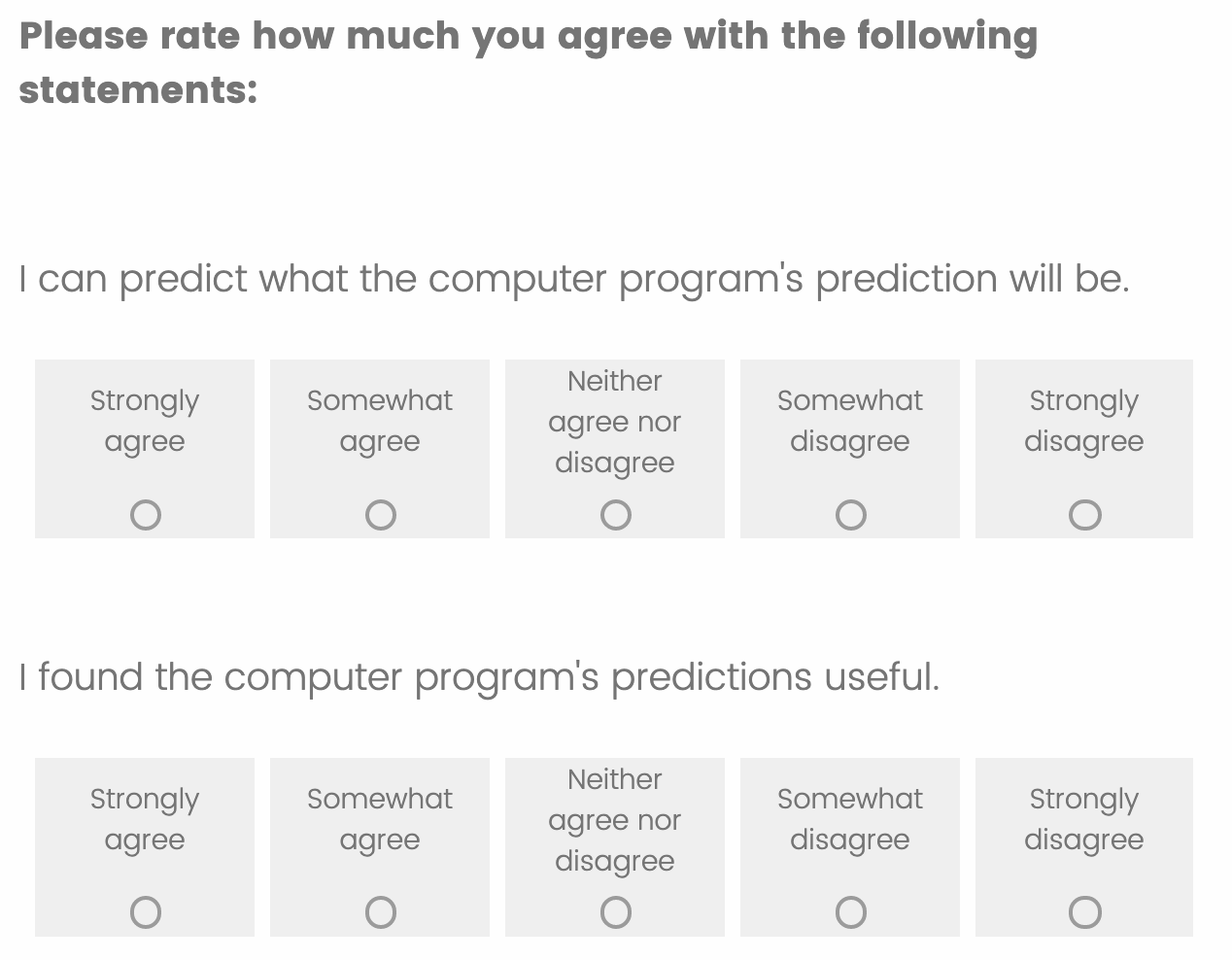}
        \caption{Assessing the predictability and usefulness of machine advice}
        \label{Fig:assess_machine}
    \end{subfigure}%
    
    \begin{subfigure}{.49\textwidth}
        \centering
        \includegraphics[width=0.99\textwidth]{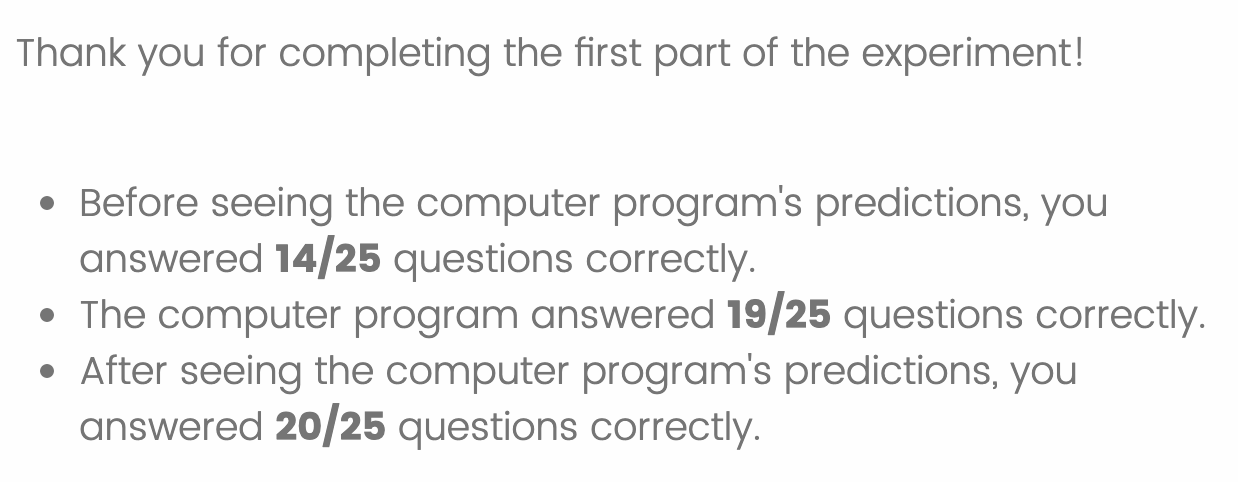}
        \caption{Feedback provided to participants after they respond to the questions shown in Figure \ref{Fig:estimate_correctness} and \ref{Fig:assess_machine}}
        \label{Fig:performance_feedback}
    \end{subfigure}%
    \caption{Eliciting participants' perceptions about the decision aid's performance}
    \label{Fig:feedback}
\end{figure*}

\begin{figure*}[t]
    \centering
    \begin{subfigure}{.45\textwidth}
        \centering
        \includegraphics[width=0.9\columnwidth]{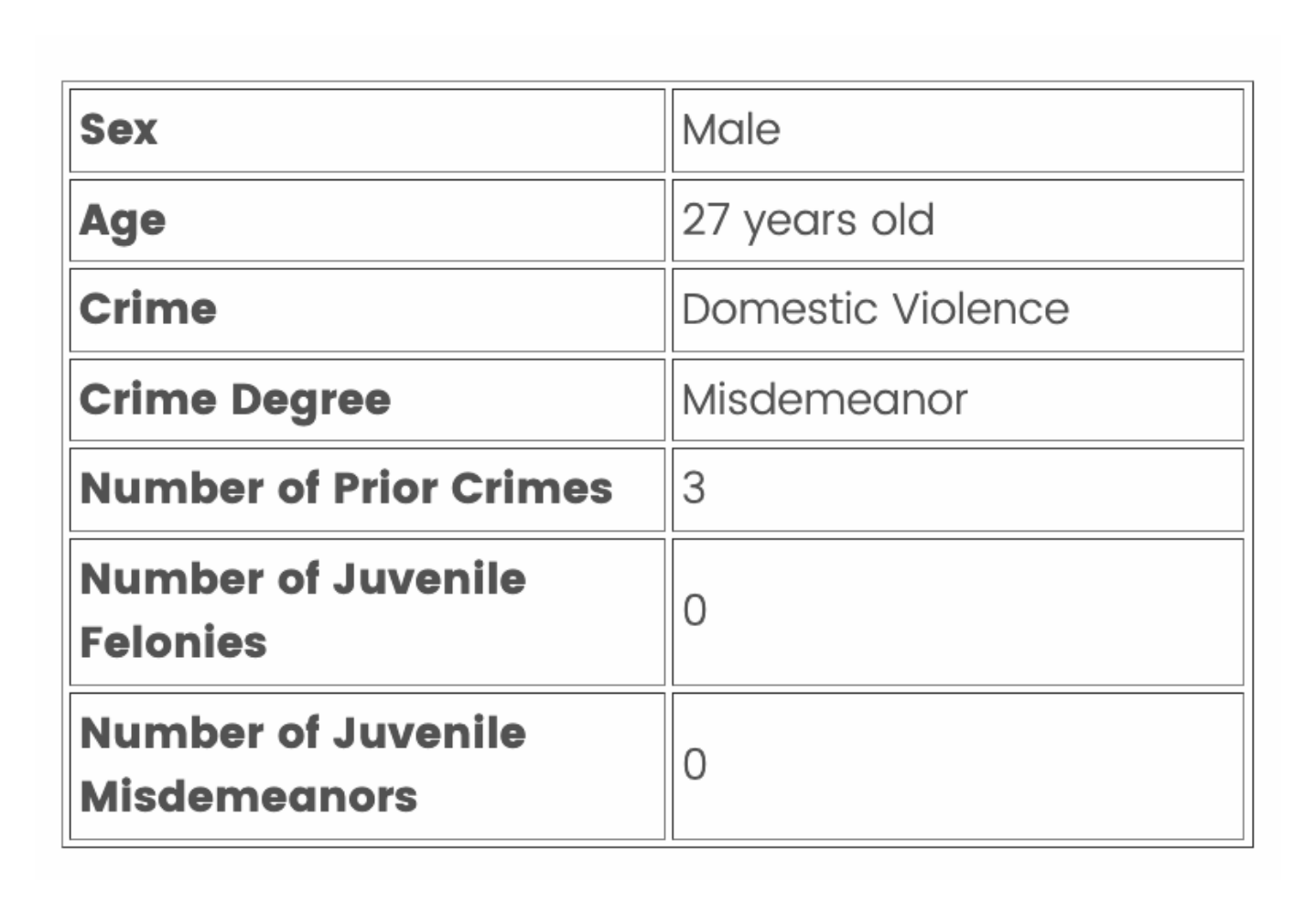}
        \caption{Defendant A}
        \label{Fig:example_human-like}
    \end{subfigure}%
    \hfill
    \begin{subfigure}{.45\textwidth}
        \centering
        \includegraphics[width=0.9\columnwidth]{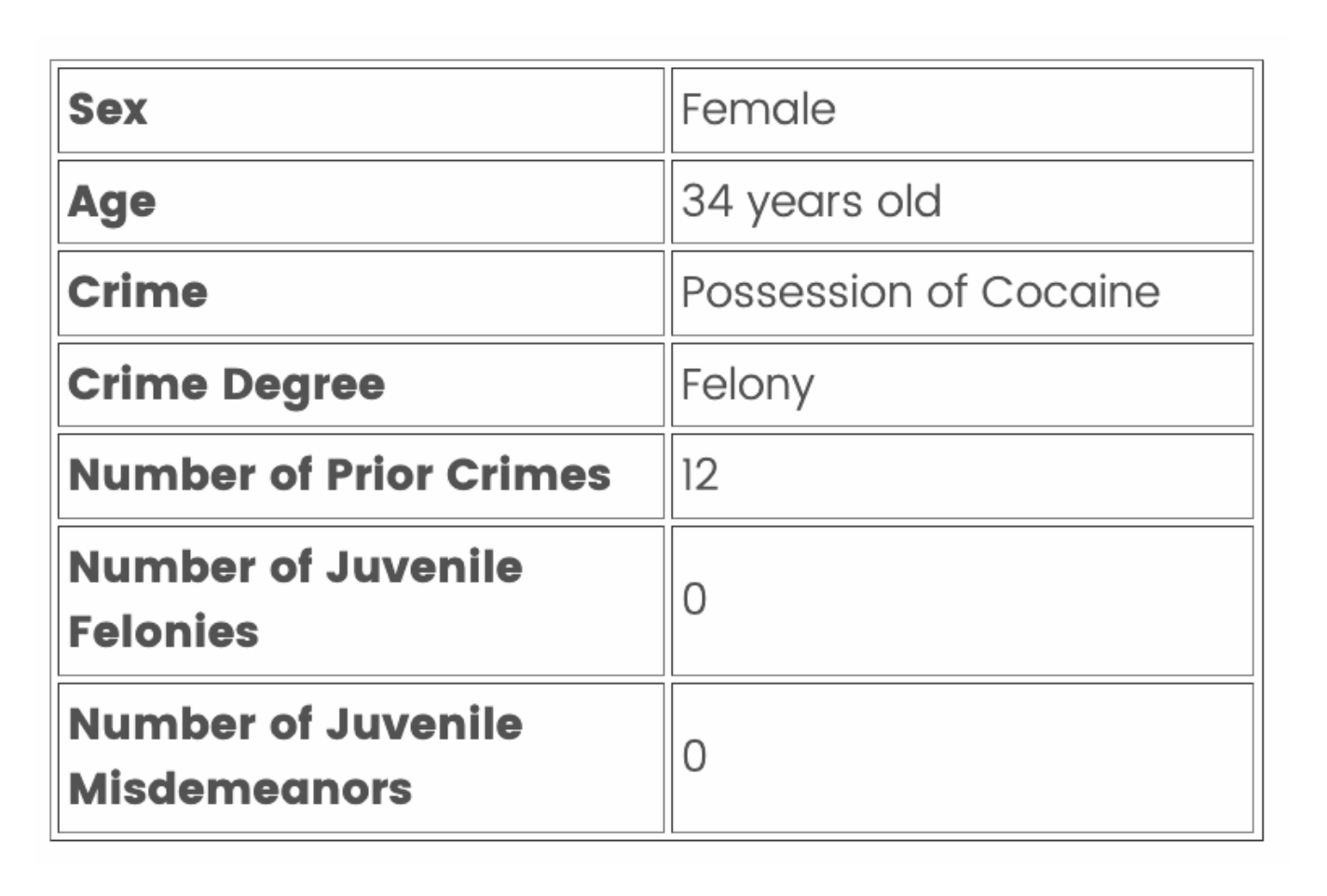}
        \caption{Defendant B}
        \label{Fig:example_anti_human-like}
    \end{subfigure}%
    \caption{Examples of mistakes made by human-like and anti human-like decision aids on the COMPAS dataset. Consider the following two defendant profiles from the COMPAS dataset. The overwhelming majority of respondents (89\% for defendant A and 94\% for defendant B) and the human-like decision aid predicted that both defendants will recidivate within two years. On the other hand, the anti human-like decision aid predicted that neither will recidivate within two years. In reality, defendant B recidivated, but defendant A did not. Hence, the human-like decision aid made a human-like error by misclassifying defendant A, while the anti human-like decision aid made a mistake few humans would make by misclassifying B.}
    \label{Fig:example_mistakes}
\end{figure*}

\xhdr{Survey Screenshots} \label{subsec:survey_screenshots}
Below, we provide screenshots of our survey instruments. 
Figure \ref{Fig:survey_intro} shows the introductory text shown to participants who were assigned to the Speed Date, High Accuracy condition. The text was identical across all treatments, except the description of the decision-making task, which varied across datasets, and the stated accuracy, which also varied across accuracy conditions.
Figure \ref{Fig:survey_questions} shows the vignettes (Fig. \ref{Fig:date_description}) and machine advice accompanied with response options (Fig. \ref{Fig:date_questions}) for the Speed Date dataset. The vignettes for the COMPAS dataset and Age dataset are shown in Figures \ref{Fig:defendant_description} and \ref{Fig:age_picture} respectively.
In Figure \ref{Fig:feedback}, we show the questions for eliciting participants' perceptions about the decision aid's performance.
Finally, Figure \ref{Fig:example_mistakes} shows an example of mistakes made by human-like and anti human-like decision aids on the COMPAS dataset.

\subsection{Additional Results} \label{subsec:additional_results}

Below, we present additional results which were omitted from the main paper due to size constraints. We report additional figures which complement the findings presented in Sections \ref{subsec:perceptions} and \ref{subsec:advice_taking}.

\begin{figure*}[t]
    \centering
    \begin{subfigure}{.33\textwidth}
        \centering
        \includegraphics[width=0.99\textwidth]{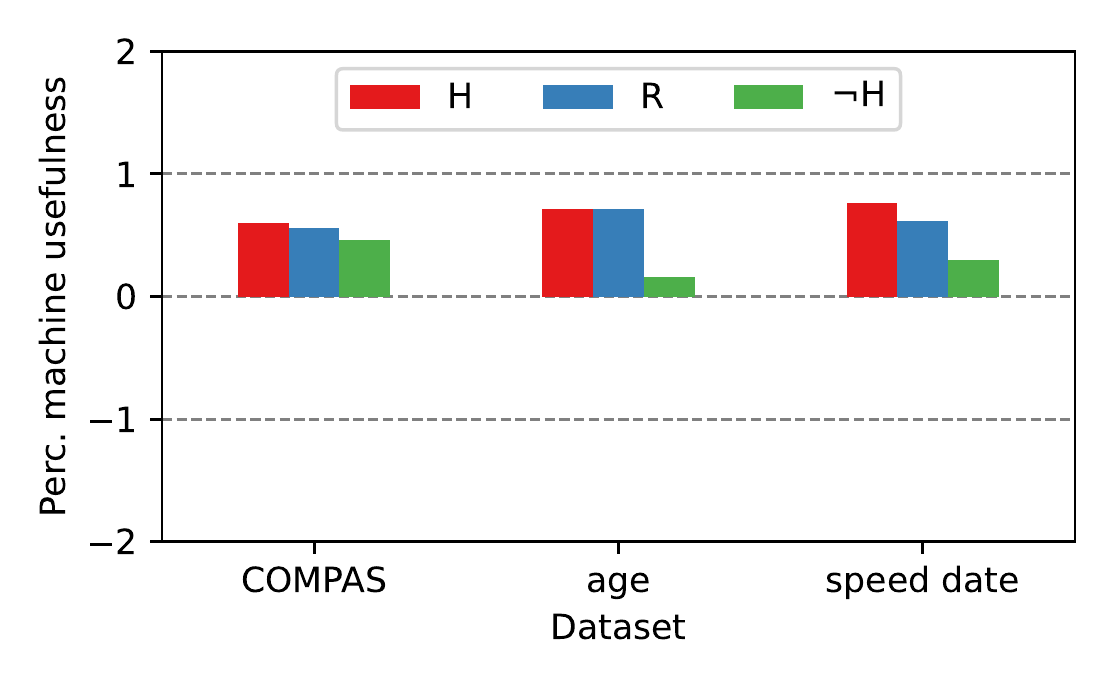} 
        \includegraphics[width=0.99\textwidth]{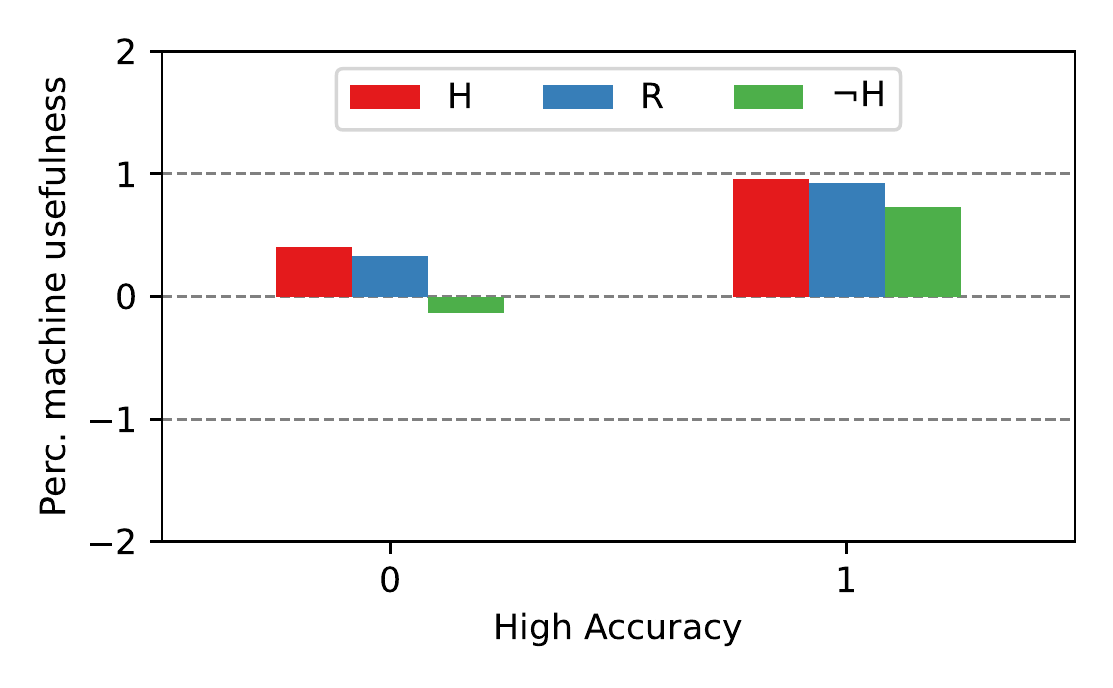} 
        \includegraphics[width=0.99\textwidth]{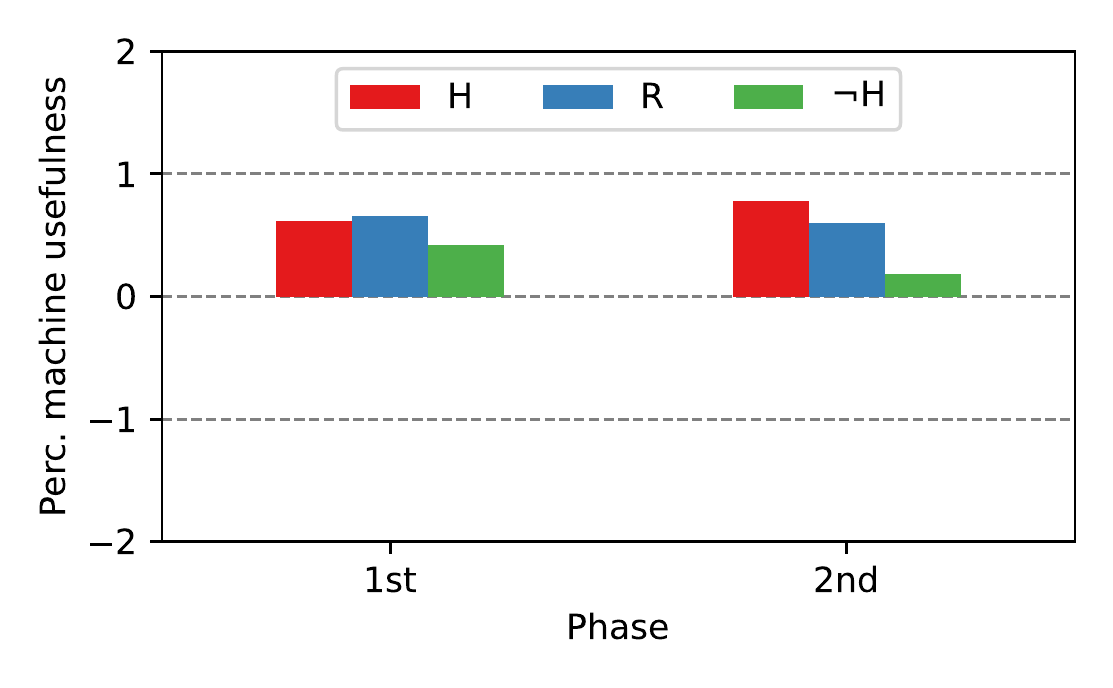}
        \caption{Usefulness}
        \label{Fig:perc_usefulness_separated}
    \end{subfigure}%
    \begin{subfigure}{.33\textwidth}
        \centering
        \includegraphics[width=0.99\textwidth]{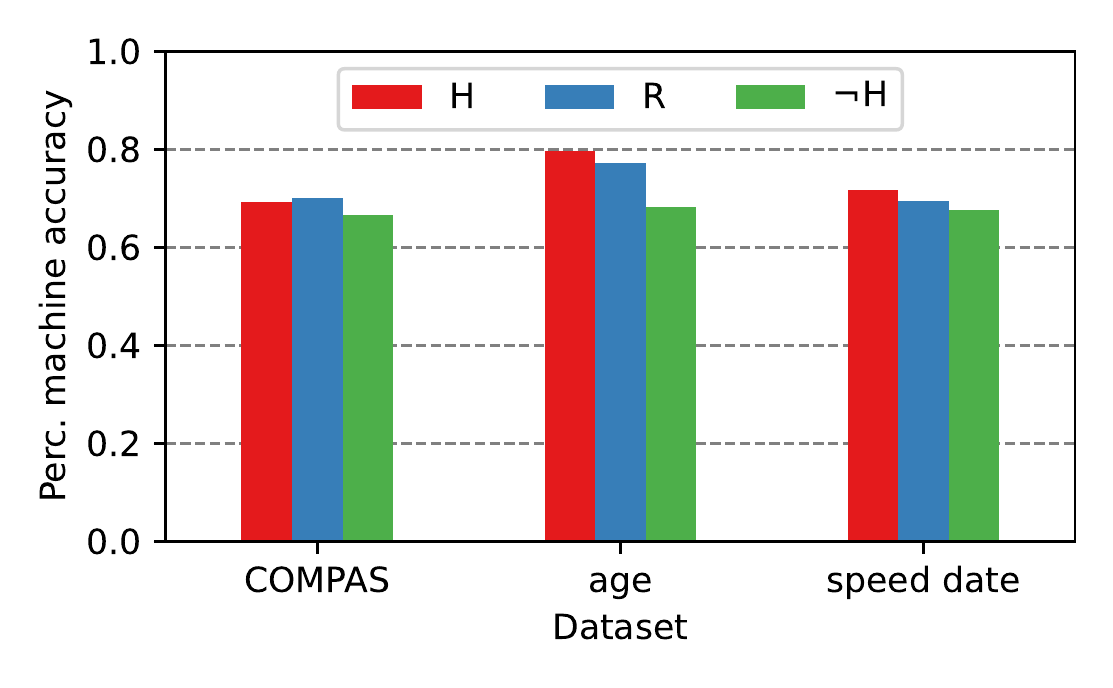}
        \includegraphics[width=0.99\textwidth]{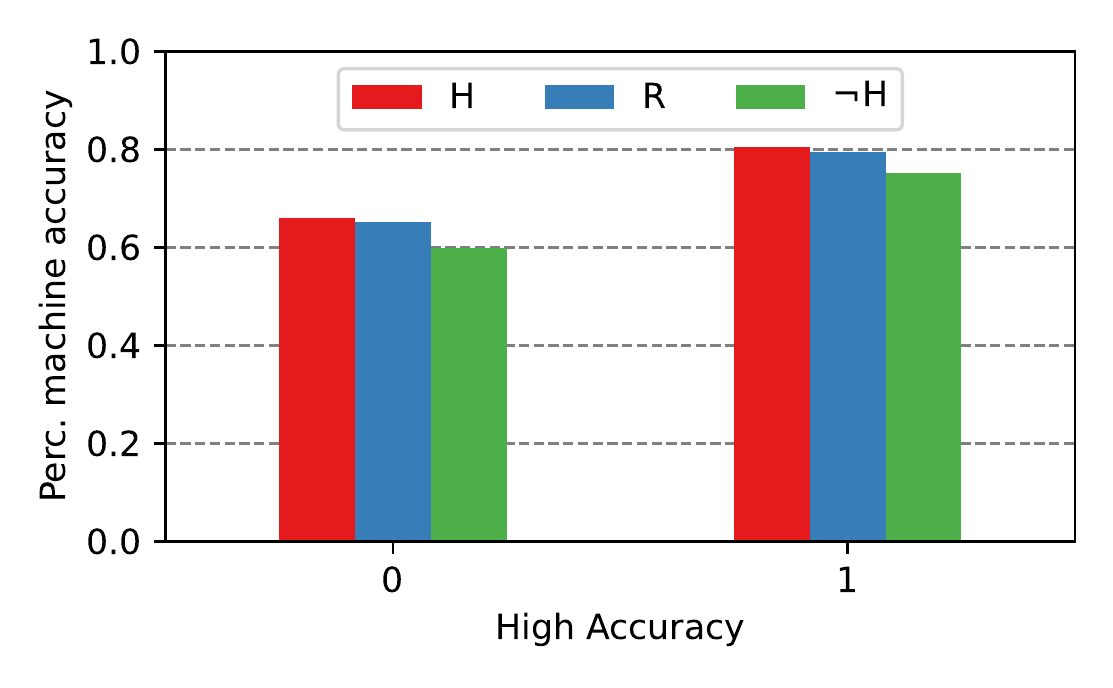}
        \includegraphics[width=0.99\textwidth]{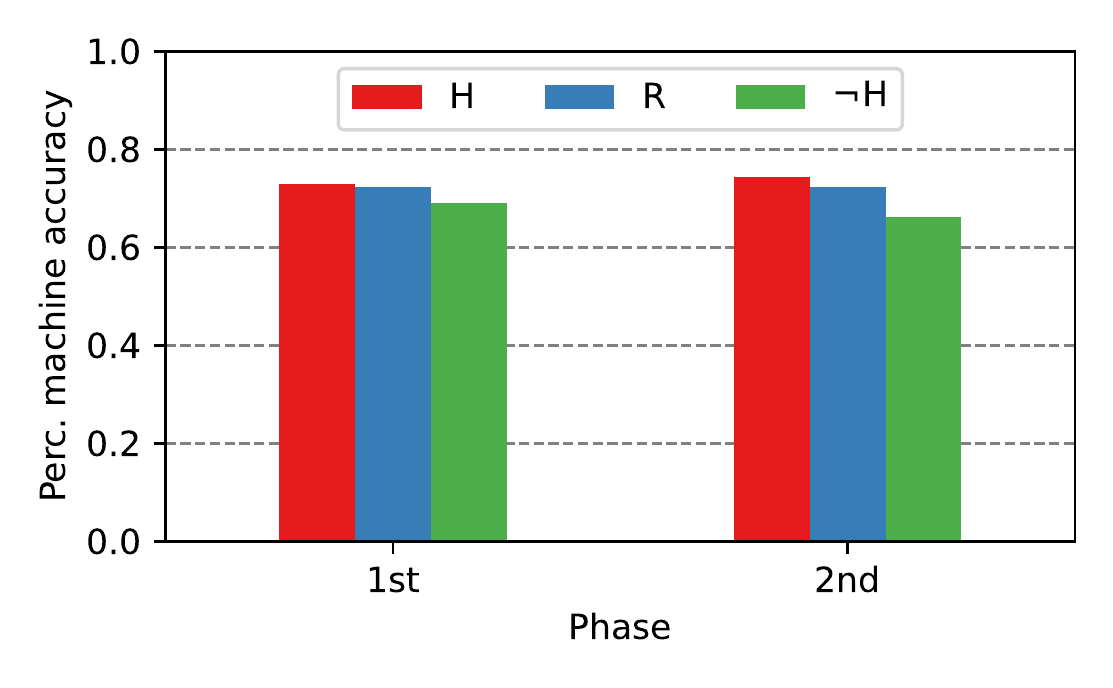}
        \caption{Accuracy}
        \label{Fig:perc_accuracy_separated}
    \end{subfigure}%
    \begin{subfigure}{.33\textwidth}
        \centering
        \includegraphics[width=0.99\textwidth]{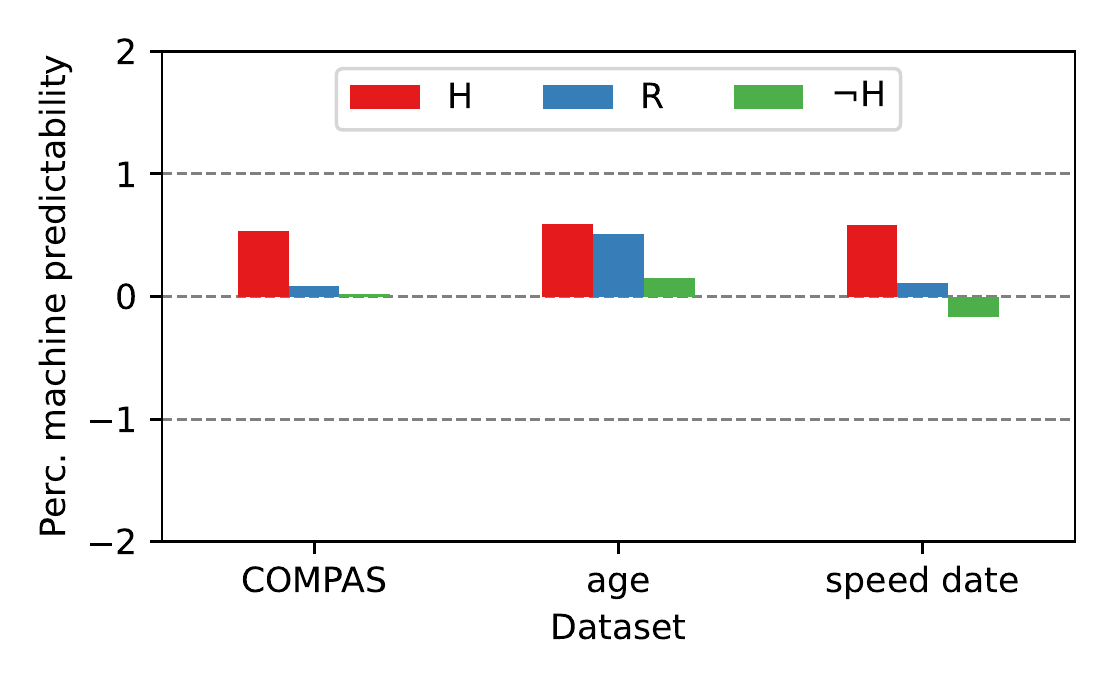}
        \includegraphics[width=0.99\textwidth]{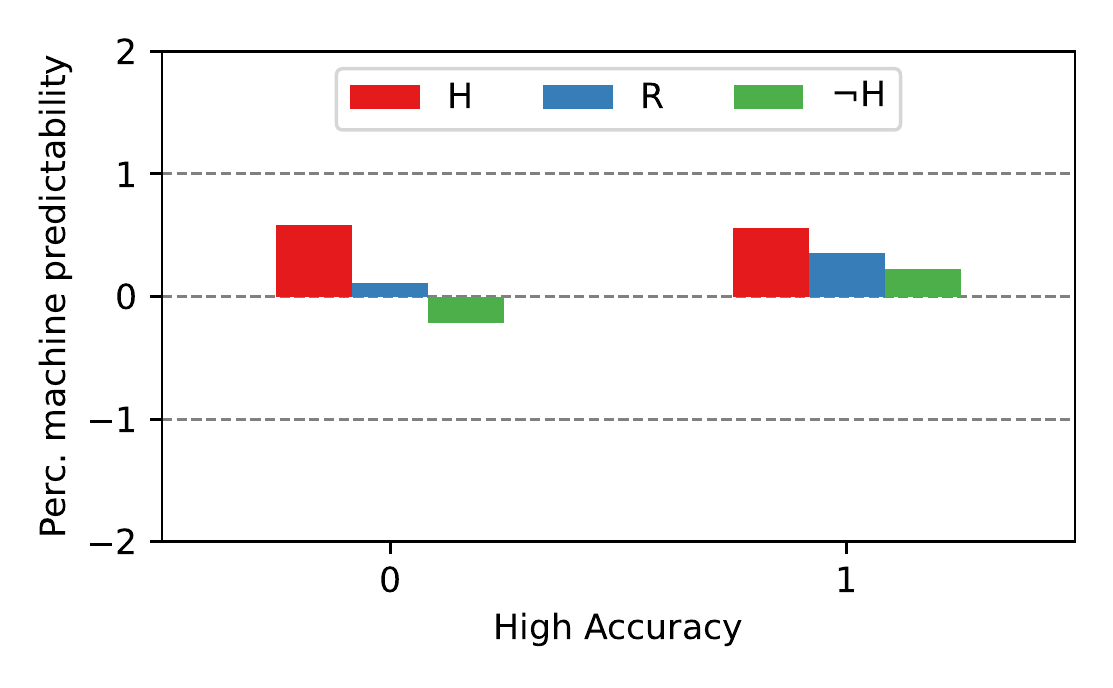}
        \includegraphics[width=0.99\textwidth]{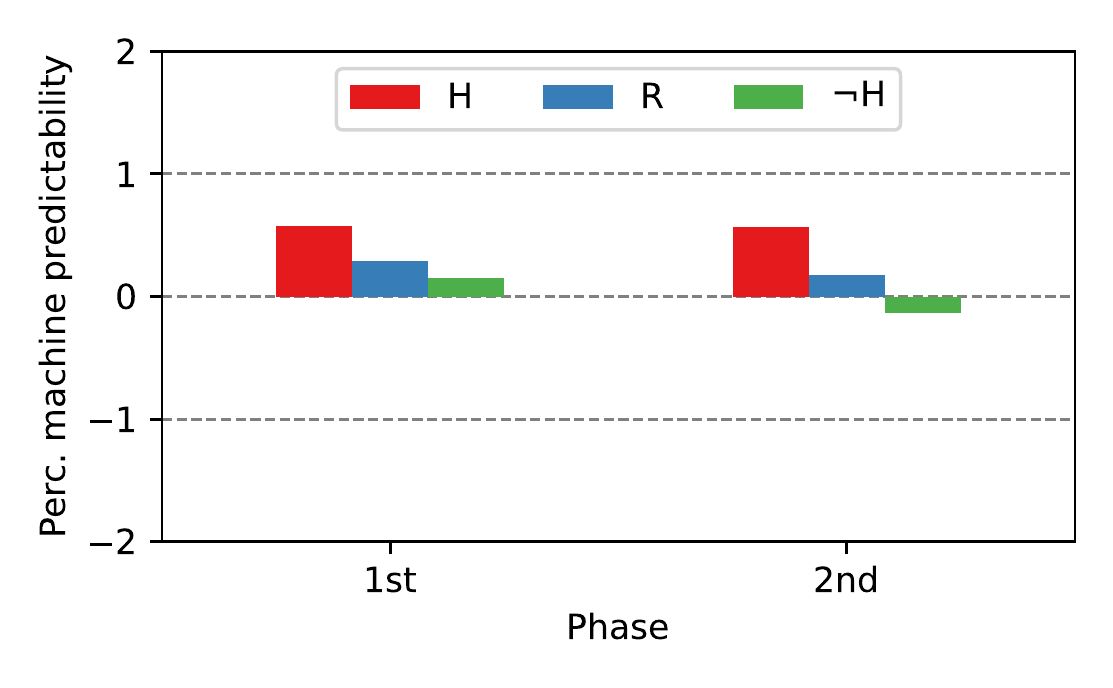}
        \caption{Predictability}
        \label{Fig:perc_predictability_separated}
    \end{subfigure}%
    \caption{Perceptions of machine predictions for decision aids with human-like (H), anti human-like ($\neg$H), and randomly distributed (R) errors. The perceptions are separated by dataset [top], accuracy [middle], and phase [bottom]. Human-like decision aids are perceived as more useful, accurate, and predictable than anti human-like decision aids across all three control variables.}
    \label{Fig:perceptions_separated}
\end{figure*}

\xhdr{Companion to Section \ref{subsec:perceptions}}
Figure \ref{Fig:perceptions_separated} complements Figure \ref{Fig:perceptions} from Subsection \ref{subsec:perceptions}. It depicts that our findings about people's perceptions of usefulness, accuracy, and predictability are robust with respect to three control variables: the dataset, the decision aid's accuracy, and the experimental phase.

\begin{figure*}[t]
    \centering
    \begin{subfigure}{.24\textwidth}
        \centering
        \includegraphics[width=0.99\textwidth]{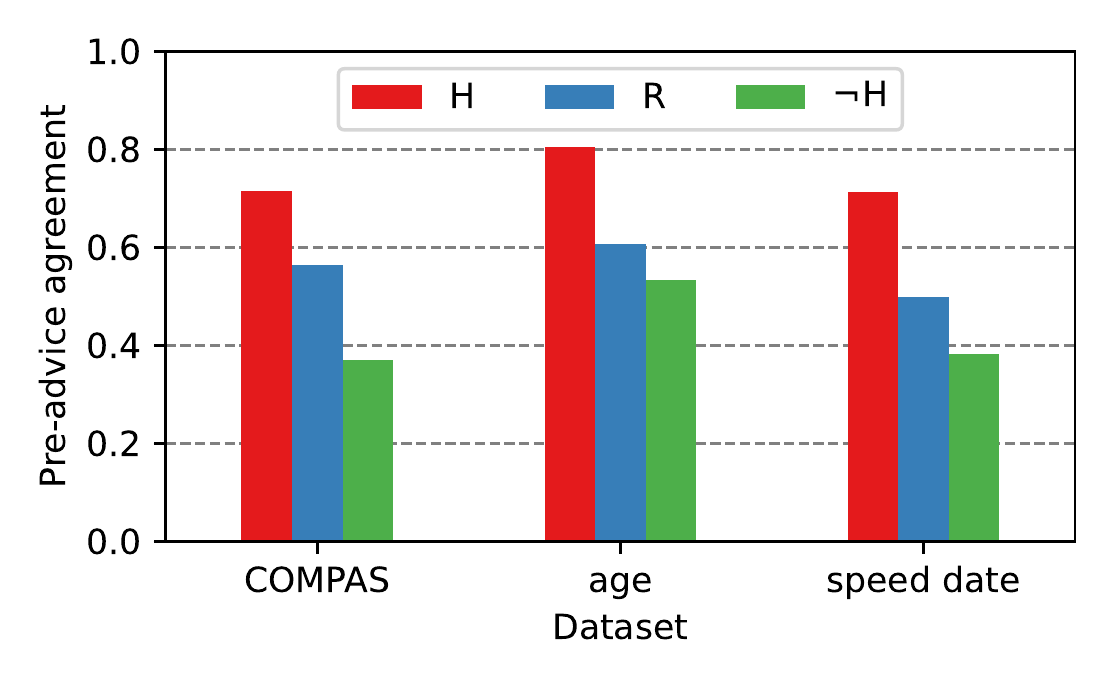}
        \includegraphics[width=0.99\textwidth]{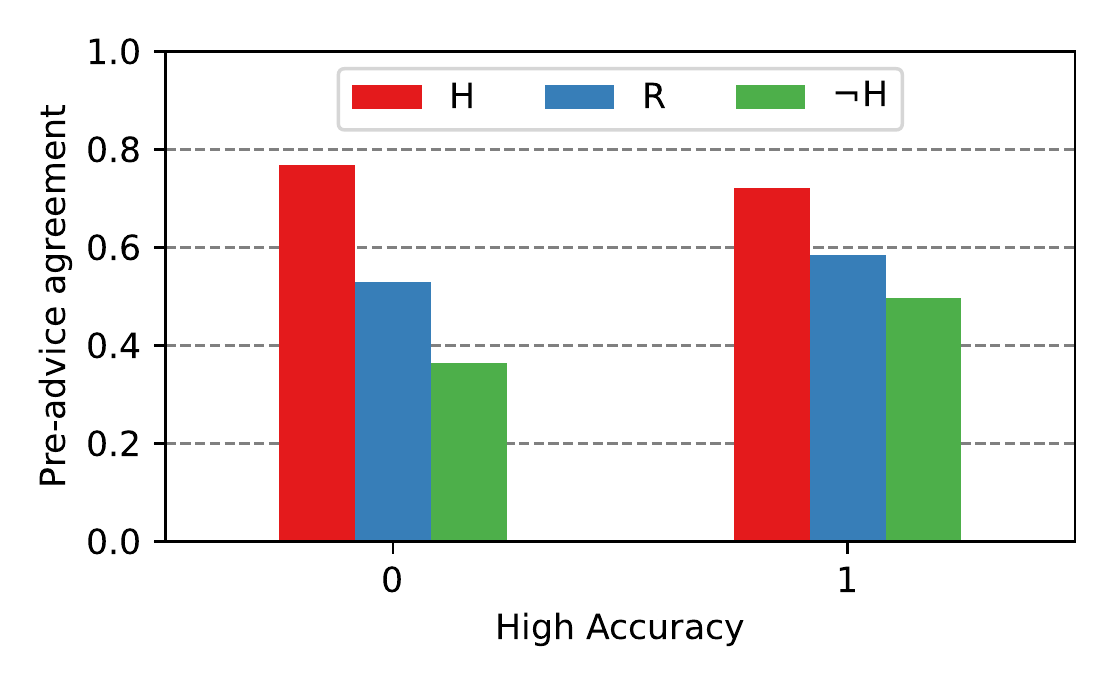}
        \includegraphics[width=0.99\textwidth]{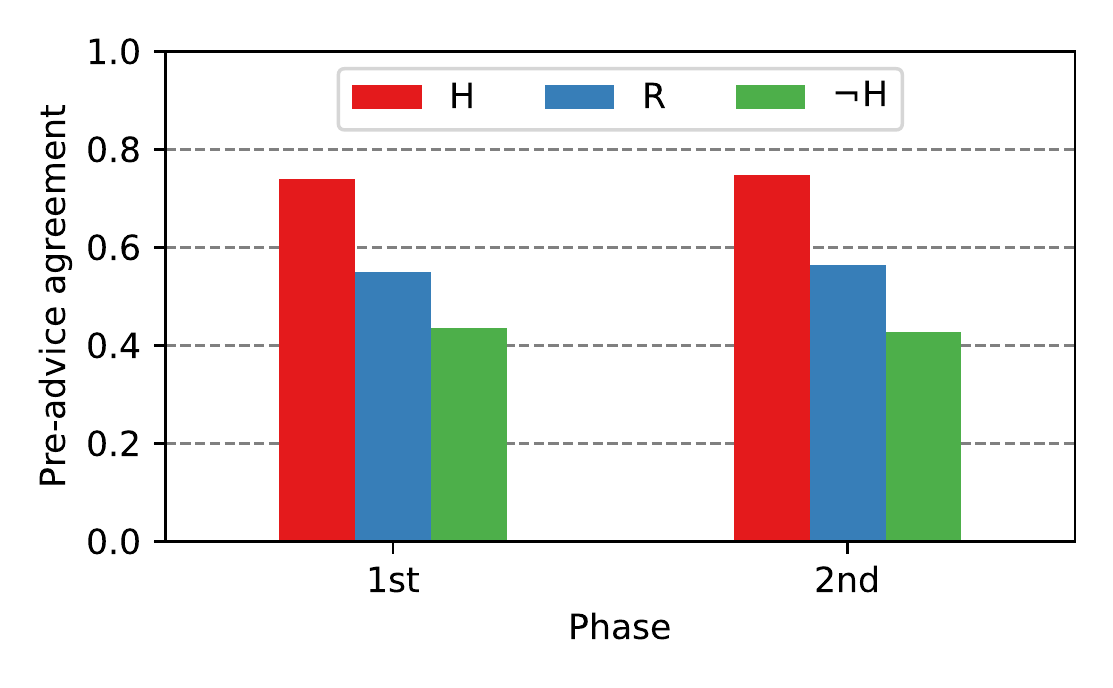}
        \caption{Pre-advice Agr.}
        \label{Fig:agreement_pre_separated}
    \end{subfigure}%
    \begin{subfigure}{.24\textwidth}
        \centering
        \includegraphics[width=0.99\textwidth]{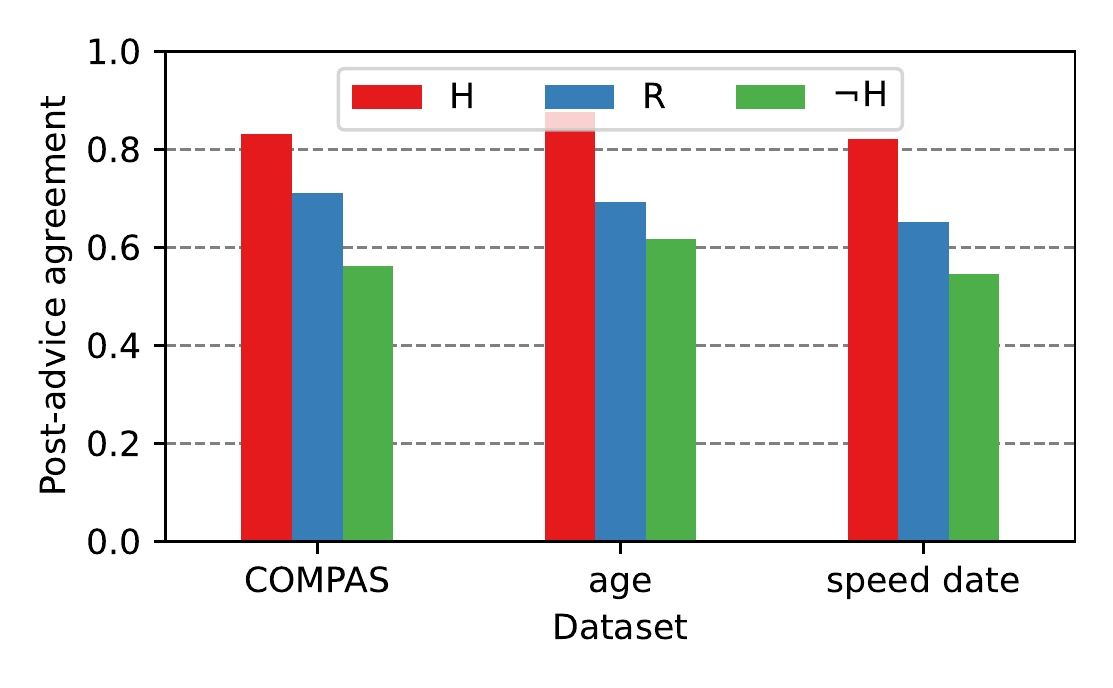}
        \includegraphics[width=0.99\textwidth]{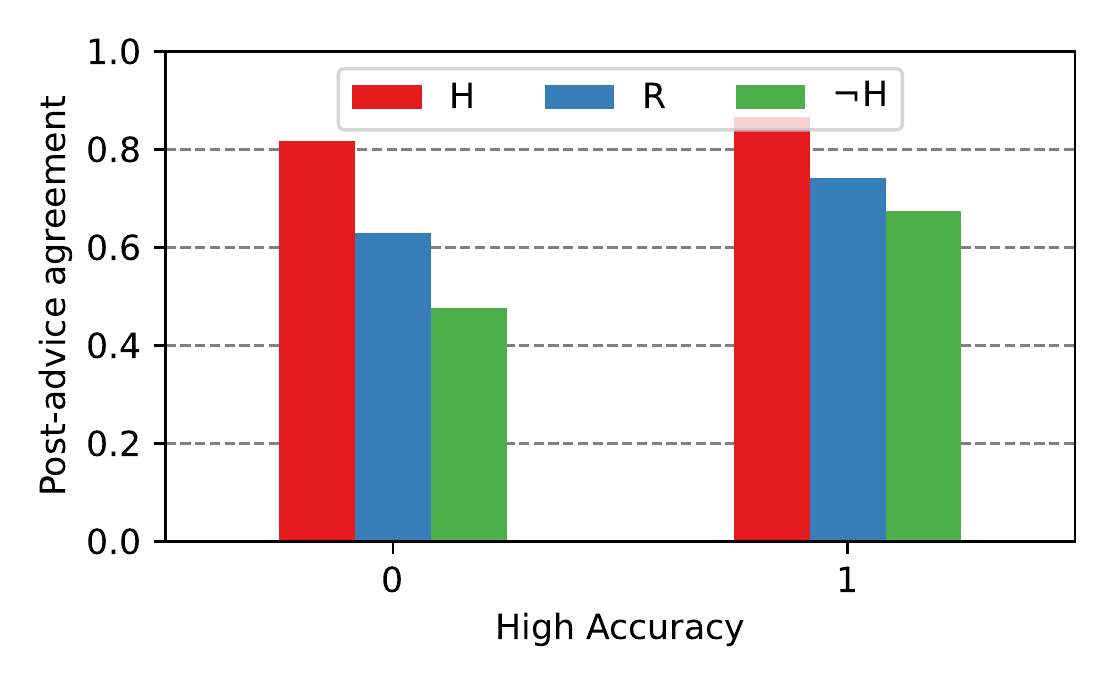}
        \includegraphics[width=0.99\textwidth]{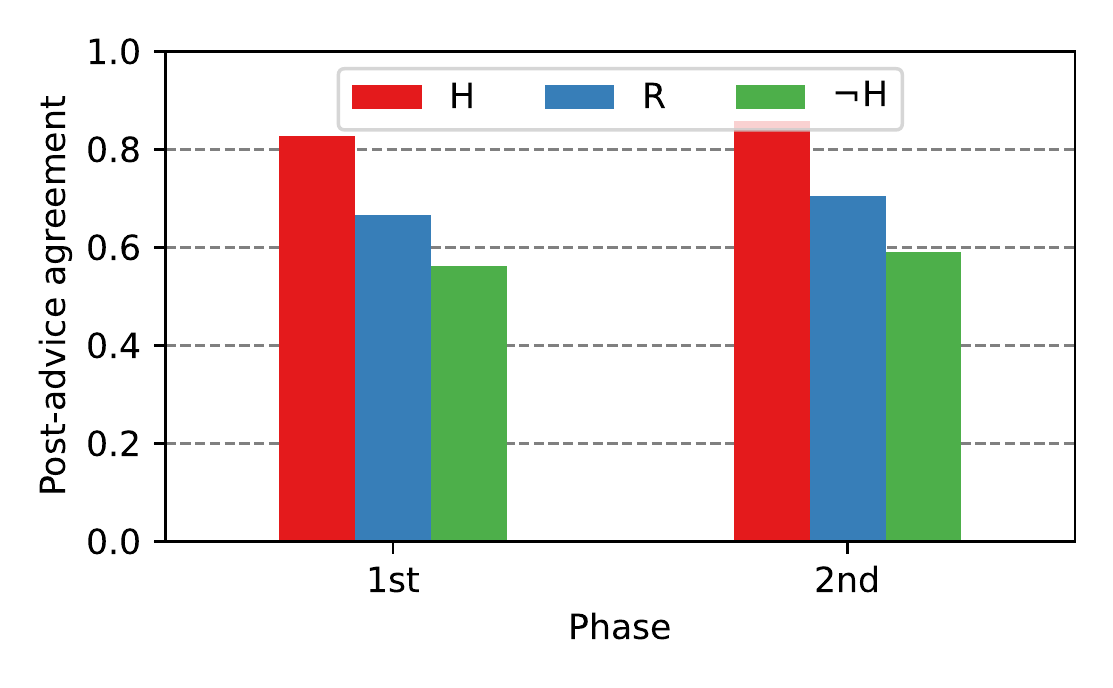}
        \caption{Post-advice Agr.}
        \label{Fig:agreement_post_separated}
    \end{subfigure}%
    \begin{subfigure}{.24\textwidth}
        \centering
        \includegraphics[width=0.99\textwidth]{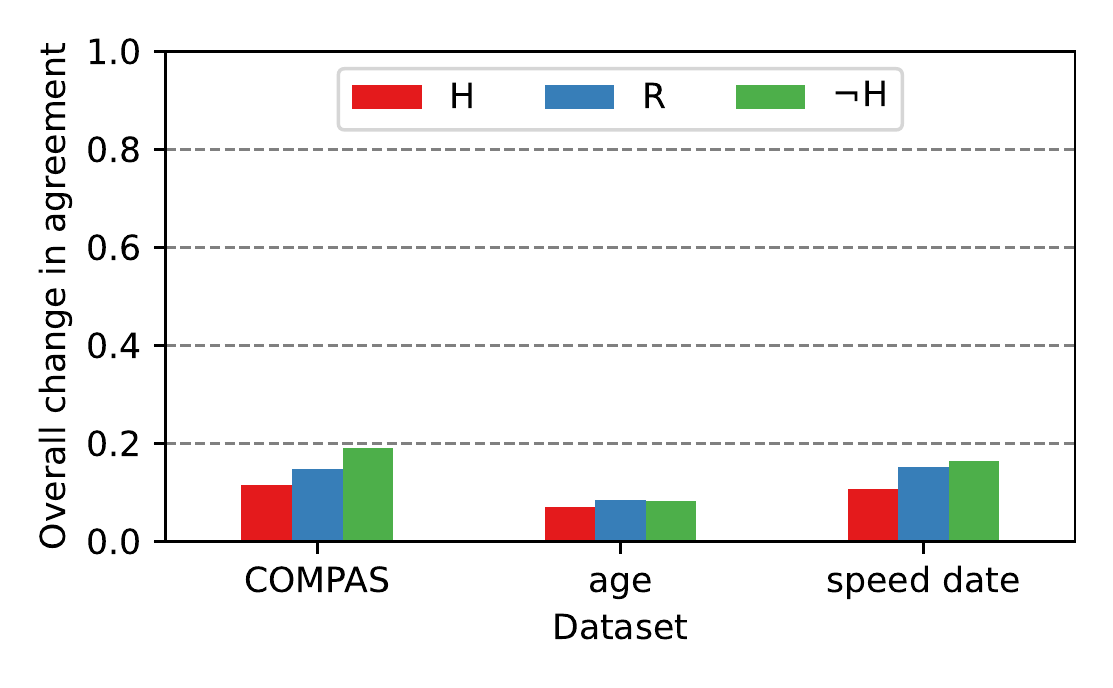}
        \includegraphics[width=0.99\textwidth]{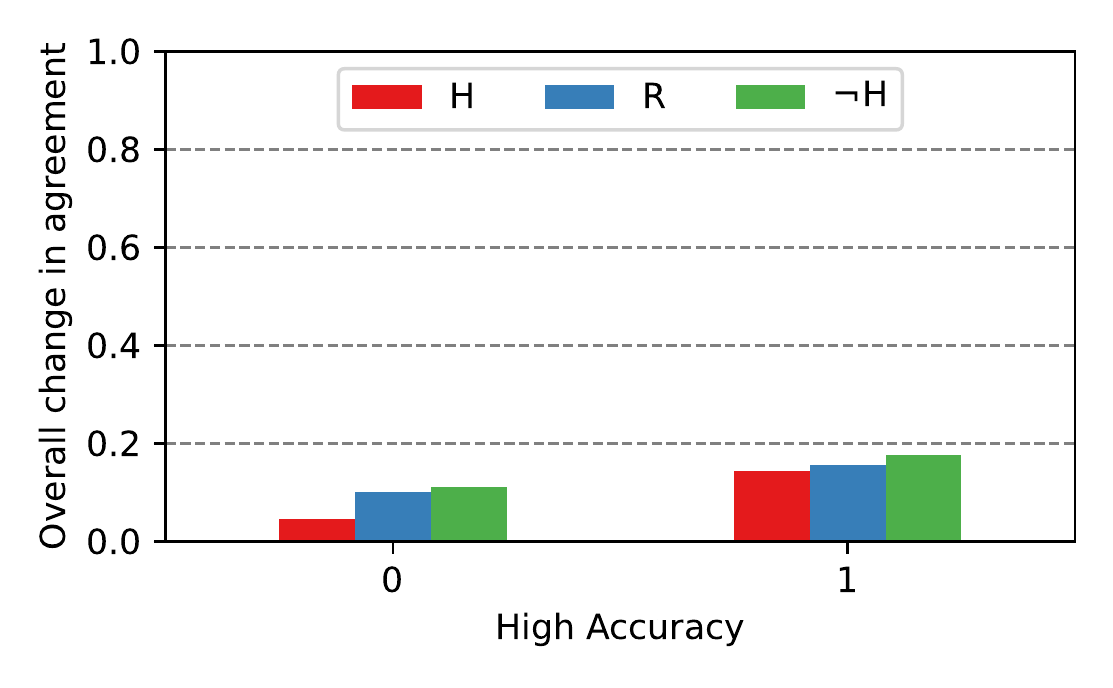}
        \includegraphics[width=0.99\textwidth]{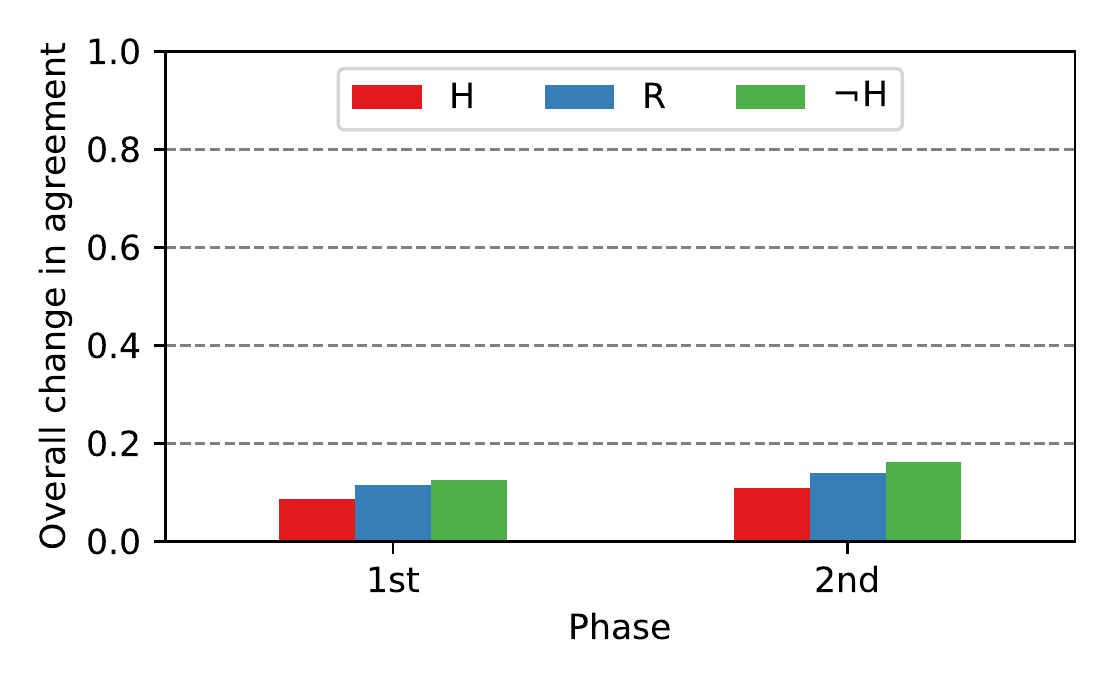}
        \caption{Overall Change in Agr.}
        \label{Fig:agreement_change_separated}
    \end{subfigure}%
    \begin{subfigure}{.24\textwidth}
        \centering
        \includegraphics[width=0.99\textwidth]{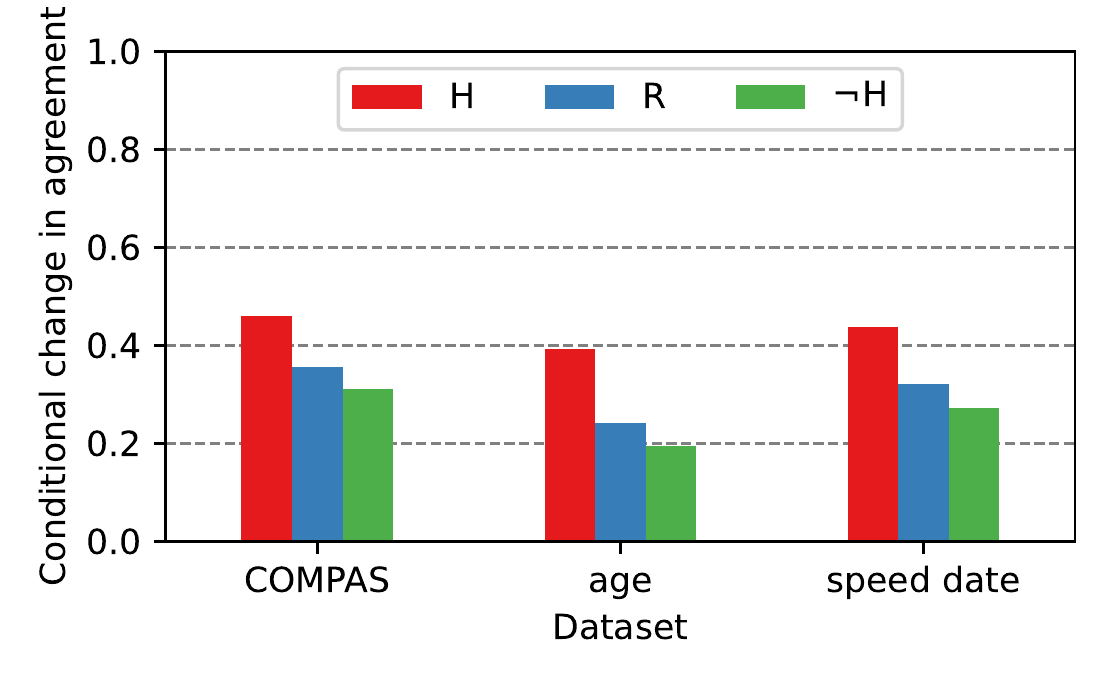}
        \includegraphics[width=0.99\textwidth]{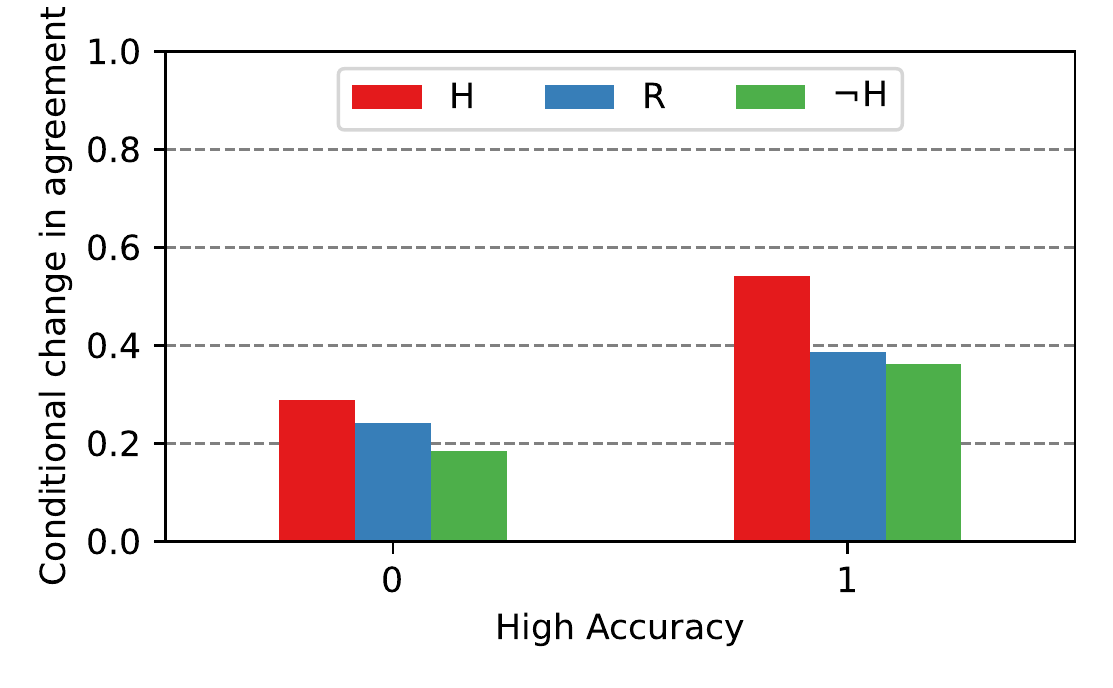}
        \includegraphics[width=0.99\textwidth]{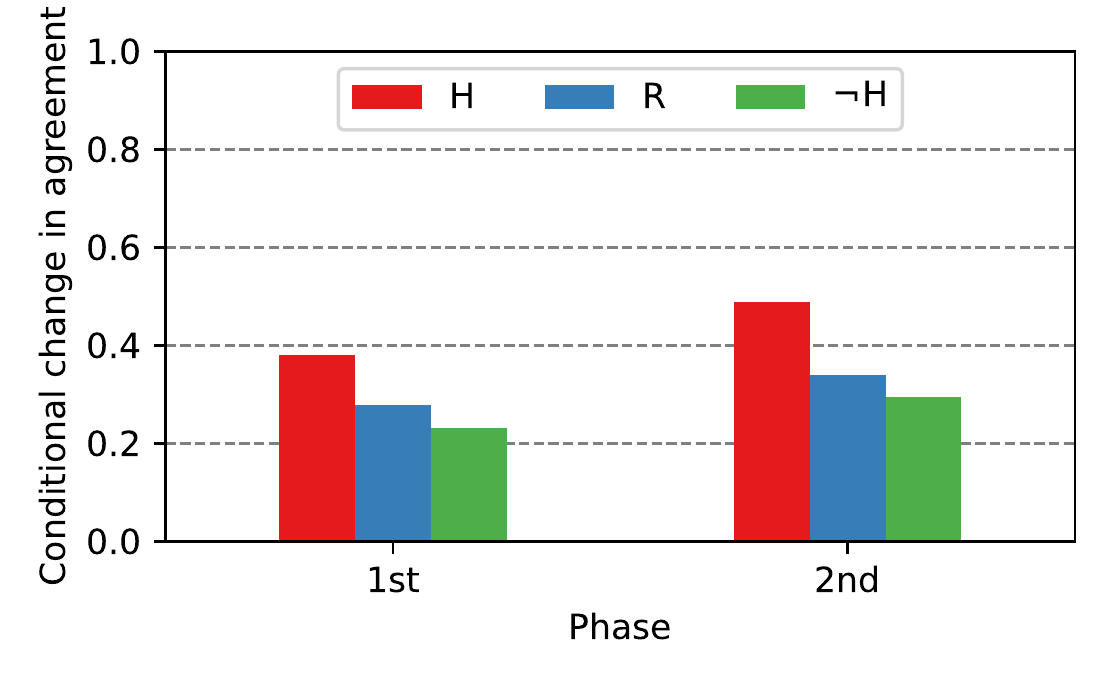}
        \caption{Cond. Change in Agr.}
        \label{Fig:agreement_change_conditional_separated}
    \end{subfigure}%
    \caption{Influence of machine advice on respondents' agreement with the advice, for decision aids with human-like (H), anti human-like ($\neg$H), and randomly distributed (R) errors. The results are separated by dataset [top], accuracy [middle], and phase [bottom]. Across all three control variables, pre and post-advice agreement, and the conditional change in agreement are higher for human-like decision aids than for anti human-like decision aids. On the other hand, for the overall agreement, the opposite pattern holds.}
    \label{Fig:agreement_separated}
\end{figure*}

\xhdr{Companion to Section \ref{subsec:advice_taking}}
Figure \ref{Fig:agreement_separated} complements Figure \ref{Fig:agreement}, showing that our findings about advice taking behavior are robust with respect to the three control variables listed above: the dataset, the decision aid's accuracy, and the experimental phase.

\end{document}